\documentclass{aa}
\usepackage[varg]{txfonts}
\usepackage{graphicx}
\usepackage{hyperref}

\newcommand{\bc}{\begin{center}}
\newcommand{\ec}{\end{center}}
\newcommand{\be}{\begin{equation}}
\newcommand{\ee}{\end{equation}}
\newcommand{\bfig}{\begin{figure}}
\newcommand{\efig}{\end{figure}}
\newcommand{\m}{\mbox}

\newcommand{\oz}{O$_3$}
\newcommand{\om}{O$_2$}
\newcommand{\od}{O($^1$D)}
\newcommand{\ch}{CH$_4$}
\newcommand{\no}{N$_2$O}
\newcommand{\hox}{HO$_x$}
\newcommand{\nox}{NO$_x$}
\newcommand{\noo}{NO$_2$}
\newcommand{\hoo}{HO$_2$}
\newcommand{\hho}{H$_2$O}

\begin{document}

\title{Is ozone a reliable proxy for molecular oxygen?} 
\subtitle{III. The impact of CH$_4$ on the \om-\oz\ relationship for Earth-like atmospheres}

\author{Thea Kozakis \inst{1,2}
  \and Jo\~ao M.\ Mendon\c{c}a \inst{2,3,4} 
  \and Lars A.\ Buchhave \inst{2}
  \and Luisa M.\ Lara \inst{1}}
  
\institute{$^1$Instituto de Astrofísica de Andalucía - CSIC, Glorieta de la Astronomía s/n, 18008 Granada, Spain \\
$^2$National Space Institute, Technical University of Denmark, Elektrovej, DK-2800 Kgs. Lyngby, Denmark\\
$^3$School of Physics and Astronomy, University of Southampton, Highfield, Southampton SO17 1BJ, UK \\
$^4$ School of Ocean and Earth Science, University of Southampton, Southampton, SO14 3ZH, UK}

\date{}							

\abstract{In the search for life in the Universe, molecular oxygen (\om) combined with a reducing species, such as methane (\ch), is considered a promising disequilibrium biosignature. In cases where it would be difficult or impossible to detect \om\ (such as in the mid-IR or low \om\ levels), it has been suggested that ozone (\oz), the photochemical product of \om, could be used as a proxy for determining the abundance of \om. As the \om-\oz\ relationship is known to be nonlinear, the goal of this series of papers is to explore how it would change for different host stars and atmospheric compositions and learning how to use \oz\ to infer \om. We used photochemistry and climate modeling to further explore the \om-\oz\ relationship by modeling Earth-like planets with the present atmospheric level (PAL) of  \om\ between 0.01\% to 150\% , along with high and low \ch\ abundances of 1000\% and 10\% PAL, respectively. Methane is of interest not only because it is a biosignature, but it is also the source of hydrogen atoms for hydrogen oxide (\hox), which destroys \oz\ through catalytic cycles, and acts as a catalyst for the smog mechanism of \oz\ formation in the lower atmosphere. We find that varying \ch\ causes changes to the \om-\oz\ relationship in ways that are highly dependent on both the host star and \om\ abundance. A striking result for high \ch\ models in high \om\ atmospheres around hotter hosts is that enough \ch\ is efficiently converted into \hho\ to significantly impact stratospheric temperatures, and therefore the formation and destruction rates of \oz. Changes in \hox\ have also been shown to influence both the \hox\ catalytic cycle and production of smog \oz, causing variations in harmful UV reaching the surface, as well as changes in the 9.6~$\mu$m \oz\ feature in emission spectra. This study further demonstrates the need to explore the \om-\oz\ relationship in different atmospheric compositions in order to use \oz\ as a reliable proxy for \om\ in future observations.
}

\keywords{astrobiology -- planets and satellites: terrestrial planets -- Planets and satellites: atmospheres}

\titlerunning{\oz\ as a proxy for molecular oxygen III}
\authorrunning{Kozakis et al.}

\maketitle

\nolinenumbers

\section{Introduction}
Although ozone (\oz) is not directly created by life, it is still included in discussions about the search for life in the Universe using atmospheric biosignatures. This is because \oz\ is the photochemical product of molecular oxygen (\om), which is primarily produced biologically on modern Earth. However, there are multiple scenarios where \om\ could build up in a planetary atmosphere in the absence of life, so \om\ on its own is not a reliable sign of life (e.g., \citealt{hu12,word14,doma14,tian14,luge15,gao15,harm15}). Instead, \om\ and a reducing species such as methane (\ch) is thought to be a promising disequilibrium biosignature, as it would require strong surface fluxes of each species, which would be indicative of surface life (e.g., \citealt{love65,lede65,lipp67}).

Ozone enters the conversation because there are scenarios in which \om\ would be difficult or impossible to detect, whereas \oz\ would be accessible. For a terrestrial planet with low levels of \om\ (such as on early Earth), it would be difficult to make an \om\ detection, while \oz\ is detectable at trace amounts \citep{kast85,lege93}. In addition, while the mid-IR (3-20~$\mu$m) holds many opportunities for biosignature searches \citep{quan22}, there are no significant \om\ features, only a collisionally induced absorption feature that is not strong enough for detections of biologically produced \om\ \citep{fauc20}. Therefore, many have suggested using \oz\ as a proxy for \om\ in such situations (e.g., \citealt{lege93,desm02,segu03,lege11,mead18,schw18}).

As \om\ and \oz\ are known to have a highly nonlinear relationship \citep{ratn72,kast80,kast85,segu03,greg21,koza22,koza25}, the goal of this series of papers is to explore the \om-\oz\ relationship for a variety of host stars and Earth-like atmospheres in order to gain valuable insight on how to use future observations of \oz\ to predict \om\ atmospheric content and potentially identify biosignatures. Here we use the term ``Earth-like'' to mean a planet with the same planetary parameters as Earth (e.g., radius, gravity), roughly the same atmospheric composition, and an orbital distance from its host star such that it receives the same total amount of flux as on modern Earth. Already we have found that not only is the \om-\oz\ relationship nonlinear, but that it varies significantly depending on the host star and planetary atmospheric composition. In \cite{koza22} we modeled planets around a variety of host stars with \om\ mixing ratios of 0.01\% to 150\% our present atmospheric level (PAL), and in \cite{koza25} we repeated those same models but with variations of high and low amounts of nitrous oxide (\no). With both studies we found that trends in the \om-\oz\ relationship differ from hotter stars to cooler stars, with the pressure-dependent nature of \oz\ formation playing a large role.

In this study we focus on the impact of \ch\ on the \om-\oz\ relationship and explore how it could impact future observations of \oz. We chose \ch\ not only because it is considered a biosignature (e.g., \citealt{thom22}), but because it is the source of hydrogen oxides (\hox) that power catalytic cycles that destroy \oz, as well as processes that create \oz\ in the lower atmosphere. Section~\ref{sec:chemistry} introduces the relevant atmospheric chemistry, Sect.~\ref{sec:methods} our methodology, and Sect.~\ref{sec:results} our results, including changes in atmospheric chemistry, UV to the ground, and \oz\ emission spectra features. In Sect.~\ref{sec:discussions} we discuss the implications of our results, and Sect.~\ref{sec:conclusions} provides our main conclusions.

\section{Relevant chemistry \label{sec:chemistry}}

\subsection{Ozone formation and destruction \label{sec:oz_formation}}

The majority of \oz\ in the atmosphere of modern Earth is formed in the stratosphere via the Chapman mechanism \citep{chap30}, beginning with \om\ photolysis,
\begin{equation}
    \m{O}_2 + \m{h}\nu \rightarrow \m{O} + \m{O  (}175 < \lambda < 242\ \m{nm}),
    \label{r:PO2_O}
\end{equation}
\vspace{-0.7cm}
\begin{equation}
    \m{O}_2 + \m{h}\nu \rightarrow \m{O }+ \m{O(}^1\m{D)  (}\lambda < 175\ \m{nm}),
    \label{r:PO2_O1D}
\end{equation}
which creates either a ground state O atom, or an excited \od\ radical depending on the energy of the photon. An \od\ radical can either be quenched back to the ground state with the help of a background molecule, $M$,
\begin{equation}
    \m{O(}^1\m{D)} + M \rightarrow \m{O} + M,
    \label{r:quench}
\end{equation}
or react with other species. Oxygen atoms can then combine with \om\ to create \oz,
\begin{equation}
    \m{O + O}_2 + M \rightarrow \m{O}_3 + M.
    \label{r:O2M}
\end{equation}
Reaction~\ref{r:O2M} requires a background molecule to carry away excess energy, meaning that it is a three-body reaction that favors higher pressures. This reaction in particular additionally favors cooler temperatures. Once \oz\ has been created, it can be photolyzed to produce \om\ and either ground state O atoms or \od\ radicals,
\begin{equation}
    \m{O}_3 + \m{h}\nu \rightarrow \m{O}_2 + \m{O(}^1\m{D)}  (\lambda < \m{310\ nm}),
    \label{r:PO3_O1D}
\end{equation}
\vspace{-0.7cm}
\begin{equation}
    \m{O}_3 + \m{h}\nu \rightarrow \m{O}_2 + \m{O (310} < \lambda < \m{1140\ nm}).
    \label{r:PO3_O}
\end{equation}
However, photolysis is not considered a real ``loss'' of \oz, as the O atom and \om\ molecule created by \oz\ photolysis often re-combine back into \oz\ (Reaction~\ref{r:O2M}). Because \oz\ and O cycle back and forth through photolysis and recombination with \om, it is useful to keep track of ``O + \oz'', called ``odd oxygen.'' A true loss of \oz\ occurs when odd oxygen is converted into \om,
\begin{equation}
    \m{O}_3 + \m{O} \rightarrow 2\m{O}_2,
    \label{r:O3_O}
\end{equation}
as \om\ photolysis is the limiting reaction of the Chapman mechanism. However, Reaction~\ref{r:O3_O} is not fast, so odd oxygen tends to be destroyed by other methods.

While the Chapman mechanism is the dominant \oz\ formation mechanism in the stratosphere, ``smog formation'' \citep{haag52} can create \oz\ in the lower atmosphere,
\begin{equation}
    \m{OH} + \m{CO} \rightarrow \m{H} + \m{CO}_2,
    \label{r:OH_CO}
\end{equation}
\vspace{-0.7cm}
\begin{equation}
    \m{H} + \m{O}_2 + M \rightarrow \m{HO}_2 + M,
    \label{r:H_O2}
\end{equation}
\vspace{-0.7cm}
\begin{equation}
    \m{HO}_2 + \m{NO} \rightarrow \m{OH} + \m{NO}_2,
    \label{r:HO2_NO}
\end{equation}
\vspace{-0.7cm}
\begin{equation}
    \m{NO}_2 + \m{h}\nu \rightarrow \m{NO} + \m{O},
    \label{r:PNO2}
\end{equation}
\vspace{-0.7cm}
\begin{equation}
    \m{O + O}_2 + M \rightarrow \m{O}_3 + M.
    \tag{\ref{r:O2M}}
\end{equation}
\begin{equation*}
\begin{aligned}
\hline
\m{Net:} \hspace{0.5cm}  \m{CO} + 2\m{O}_2 + \m{h}\nu \rightarrow \m{CO}_2 + \m{O}_3
\end{aligned}
\end{equation*}
The smog mechanism requires both nitrogen oxides (\nox, NO$_3$+\noo+NO) and hydrogen oxides (\hox, \hoo+OH+H) as catalysts to form \oz; neither are consumed in the process.

A true loss of odd oxygen/\oz\ most typically occurs from catalytic cycles, which take the form,
\begin{equation*}
\begin{aligned}
\m{X + O}_3 \rightarrow \m{XO + O}_2, \\
\m{XO + O} \rightarrow \m{X + O}_2, \\
\hline
\m{Net:} \hspace{0.5cm}  \m{O}_3 + \m{O} \rightarrow 2\m{O}_2
\end{aligned}
\end{equation*}
with the primary catalytic cycles on Earth being that of the \nox\ (X=NO) and \hox\ (X=OH) catalytic cycles. The main \nox\ catalytic cycle is,
\begin{equation}
\m{NO} + \m{O}_3 \rightarrow \m{NO}_2 + \m{O}_2,
\label{r:NO_O3}
\end{equation}
\vspace{-0.7cm}
\begin{equation}
\m{NO}_2 + \m{O} \rightarrow \m{NO} + \m{O}_2,
\end{equation}
with a secondary \nox\ catalytic cycle working in the lower stratosphere using NO$_3$,
\begin{equation}
\m{NO} + \m{O}_3 \rightarrow \m{NO}_2 + \m{O}_2,
\tag{\ref{r:NO_O3}}
\end{equation}
\vspace{-0.7cm}
\begin{equation}
\m{NO}_2 + \m{O}_3 \rightarrow \m{NO}_3 + \m{O}_2.
\end{equation}
\vspace{-0.7cm}
\begin{equation}
\m{NO}_3 + \m{h}\nu \rightarrow \m{NO} + \m{O}_2.
\end{equation}
The primary mechanism for removing \nox\ from the atmosphere is by conversion into stable reservoir species, 
\begin{equation}
    \m{NO}_2 + \m{NO}_3 + M \rightarrow \m{N}_2\m{O}_5 + M,
    \label{r:NO2_NO3_N2O5}
\end{equation}
\vspace{-0.7cm}
\begin{equation}
    \m{OH} + \m{NO}_2 + M \rightarrow \m{HNO}_3 + M,
    \label{r:OH_NO2_HNO3}
\end{equation}
\vspace{-0.7cm}
\begin{equation}
    \m{HO}_2 + \m{NO}_2 + M \rightarrow \m{HO}_2\m{NO}_2 + M.
    \label{r:HO2_NO2_HO2NO2}
\end{equation}

\subsection{Methane and HO$_x$ catalytic cycles \label{sec:chemistryCH4}}
Along with the \nox\ catalytic cycle the other dominant destruction of \oz\ is via the \hox\ catalytic cycle. While H$_2$O is the direct source of \hox\ in the atmosphere, H$_2$O has limited upward transport due to the cold trap, which prevents travel from the troposphere into the stratosphere, hence the significantly drier stratosphere than the troposphere on modern Earth. However, \ch\ can freely move from the ground into the stratosphere, where it can be oxidized into H$_2$O molecules. As a result, \ch\ is the primary source of stratosphere \hox, and thus powers the \hox\ catalytic cycle.

Another popular biosignature, \ch, is produced on Earth primarily from natural wetlands, although it has strong anthropogenic sources such as rice paddies and other agricultural processes that are not addressed in this study. When a molecule of \ch\ is transported into the stratosphere, it has the potential to create two H$_2$O molecules given enough oxygen and incoming UV flux. To start off this process, \ch\ is oxidized by the hydroxyl radical (OH), to create H$_2$O and the methyl radical CH$_3$,
\begin{equation}
    \m{CH}_4 + \m{OH} \rightarrow \m{H}_2\m{O} + \m{CH}_3,
    \label{r:CH4_OH}
\end{equation}
\vspace{-0.7cm}
\begin{equation}
    \m{CH}_3 + \m{O}_2 + M \rightarrow \m{CH}_2\m{O} + \m{OH} + M,
    \label{r:CH3_O2}
\end{equation}
\vspace{-0.7cm}
\begin{equation}
    \m{CH}_2\m{O} + \m{OH} \rightarrow \m{H}_2\m{O} + \m{CHO},
    \label{r:CH2O_OH}
\end{equation}
\begin{equation*}
\begin{aligned}
\hline
\m{Net:} \hspace{0.5cm}  \m{CH}_4 + \m{OH} + \m{O}_2 \rightarrow 2\m{H}_2\m{O} + \m{HCO}
\end{aligned}
\end{equation*}
with the net result of two H$_2$O molecules. Although there is technically a loss of \hox\ (through OH), the resulting formyl radical (HCO) is frequently converted back into \hox\ via reactions with \om,
\begin{equation}
    \m{HCO} + \m{O}_2 \rightarrow \m{CO} + \m{HO}_2.
    \label{r:HCO_O2}
\end{equation}
The original source of the OH in Reaction~\ref{r:CH4_OH} is often via \od\ reacting with H$_2$O,
\begin{equation}
    \m{H}_2\m{O} + \m{O(}^1\m{D)} \rightarrow \m{OH} + \m{OH}.
    \label{r:H2O_O1D}
\end{equation}
Since \od\ is created by photolysis with high energy UV photons, production is highly dependent on the spectrum of the host star, which we explore in depth in the rest of this study.

Although on modern day Earth the biggest sink of the short-lived OH radical is \ch, OH is also the main sink of \ch\ in the atmosphere. Additionally OH is commonly created by \od\ resulting from \oz\ photolysis, making \oz\ an indirect source of \hox. This creates interesting feedback behavior, as \hox\ is a significant sink of \oz\ via the \hox\ catalytic cycle,
\begin{equation}
\m{OH} + \m{O}_3 \rightarrow \m{HO}_2 + \m{O}_2,
\label{r:HOx_OH}
\end{equation}
\vspace{-0.7cm}
\begin{equation}
\m{HO}_2 + \m{O} \rightarrow \m{OH} + \m{O}_2,
\label{r:HOx_HO2}
\end{equation}
in which odd oxygen, O + \oz, is converted into two \om\ molecules. In the upper stratosphere, where H atoms are more common (often from H$_2$O photolysis), odd oxygen can be converted to \om\ via,
\begin{equation}
\m{H} + \m{O}_2 + M \rightarrow \m{HO}_2 + M,
\label{r:HM}
\end{equation}
\vspace{-0.7cm}
\begin{equation}
\m{HO}_2 + \m{O} \rightarrow \m{OH} + \m{O}_2,
\tag{\ref{r:HOx_HO2}}
\end{equation}
\vspace{-0.7cm}
\begin{equation}
\m{OH} + \m{O} \rightarrow \m{H} + \m{O}_.
\end{equation}
In the lower stratosphere where there is less \om\ photolysis and therefore fewer O atoms, odd oxygen is destroyed via,
\begin{equation}
\m{OH} + \m{O}_3 \rightarrow \m{HO}_2 + \m{O}_2,
\end{equation}
\vspace{-0.7cm}
\begin{equation}
\m{HO}_2 + \m{O}_3 \rightarrow \m{OH} + 2\m{O}_2.
\end{equation}
There are multiple reactions that destroy either OH or \hoo, but they are typically recycled back into another \hox\ species. Photolysis is also not a true sink of \hox\ since \hoo\ photolysis creates OH, and OH is too short-lived for significant photolysis. Efficient methods of \hox\ destruction are conversion to H$_2$O,
\begin{equation}
    \m{OH} + \m{HO}_2 \rightarrow \m{H}_2\m{O}+ \m{O}_2,
    \label{r:OH_HO2_H2O}
\end{equation}
since the conversion of H$_2$O into \hox\ is a limiting reaction. \hox\ is also lost via conversion to a stable reservoir species,
\begin{equation}
    \m{OH} + \m{OH} + M \rightarrow \m{H}_2\m{O}_2 + M,
    \label{r:OH_OH_H2O2}
\end{equation}
\vspace{-0.7cm}
\begin{equation}
    \m{OH} + \m{NO}_2 + M \rightarrow \m{HNO}_3 + M,
    \tag{\ref{r:OH_NO2_HNO3}}
\end{equation}
\vspace{-0.7cm}
\begin{equation}
    \m{HO}_2 + \m{NO}_2 + M \rightarrow \m{HO}_2\m{NO}_2 + M.
   \tag{\ref{r:HO2_NO2_HO2NO2}}
\end{equation}

\begin{figure}[h!]
    \centering
    \includegraphics[scale=0.3]{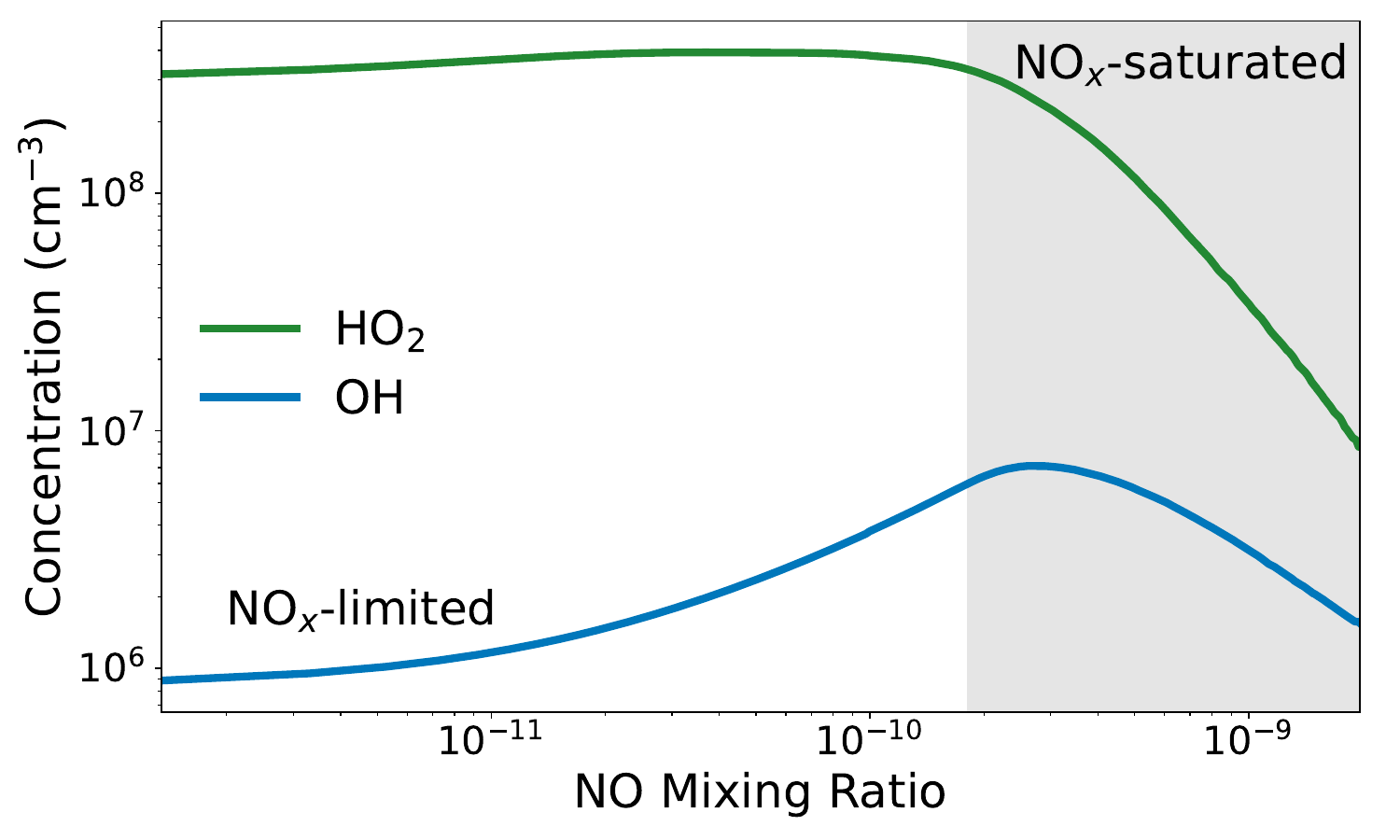}
    \caption{Relationship between \nox\ and \hox\ in the lower atmosphere on modern Earth, adapted from \cite{loga81}. In the \nox-limited regime (white background) an increase of \nox\ will lead to an increase of \hox\ and thus increased productivity of the smog mechanism. In the \nox-saturated regime (gray background) the \nox/\hox\ ratio reaches a tipping point where \nox\ begins to deplete \hox\ by locking it up into stable reservoir species, therefore suppressing smog formation.
    \label{fig:NOx_regimes}}
\end{figure}

\subsection{Relationship between NO$_x$ and HO$_x$ \label{sec:NOx_regimes}}
As mentioned previously, \oz\ is often an indirect source of \hox\ as \od\ radicals that react with \hho\ to form \hox (Reaction~\ref{r:H2O_O1D}) are typically formed via \oz\ photolysis (Reaction~\ref{r:PO3_O1D}) in \om-rich atmospheres. This causes an interesting relationship with the smog mechanism of \oz\ production (Reactions~\ref{r:O2M}, \ref{r:OH_CO}, \ref{r:H_O2}, \ref{r:HO2_NO}, \ref{r:PNO2}), as both \hox\ and \nox\ are required as catalysts. Earth-based studies have found that the amount of \nox\ present in a region can either increase the rate of smog produced \oz\ (and thus indirectly increasing \hox) or suppress \oz\ production by locking \hox\ up into reservoir species such as HNO$_3$ (Reaction~\ref{r:OH_NO2_HNO3}) or HO$_2$NO$_2$ (Reaction~\ref{r:HO2_NO2_HO2NO2}). We refer to these two scenarios in which \nox\ can assist smog production or hinder it as the ``\nox-limited'' and ``\nox-saturated'' regimes, respectively, as illustrated in Fig.~\ref{fig:NOx_regimes}. In the \nox-limited regime, increasing \nox\ leads to more \oz\ and therefore more \hox, encouraging smog formation, and in the \nox-saturated regime, the \nox/\hox\ ratio has become high enough that \nox\ locks up \hox\ in reservoir species and suppresses smog formation.

However, it is important to note that these regimes have been studied primarily in the context of \nox\ pollution on modern Earth via anthropogenic activities, where the emphasis is placed on how changes in \nox\ (not changes in \hox) impact smog \oz\ formation. This study provides a new perspective on these \nox\ regimes, as we focus on changes in \hox\ caused by varying \ch\ levels. These regimes were discussed in depth in \cite{koza25} where we varied \no\ (and therefore \nox) and we found that even for modern levels of \no\ and \om\ the \nox\ abundances were already high enough to be in the \nox-saturated regime for planets around all hosts explored in this paper except for the M5V at modern \om\ levels, causing severe suppression of the smog mechanism.

\begin{figure}[h!]
    \centering
    \includegraphics[scale=0.45]{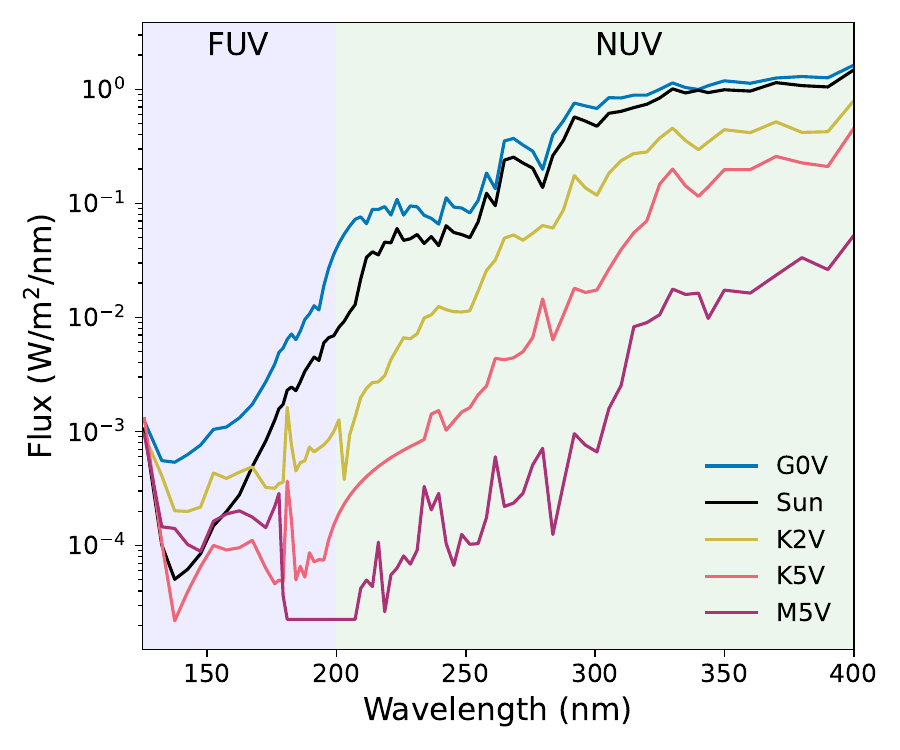}
    \includegraphics[scale=0.45]{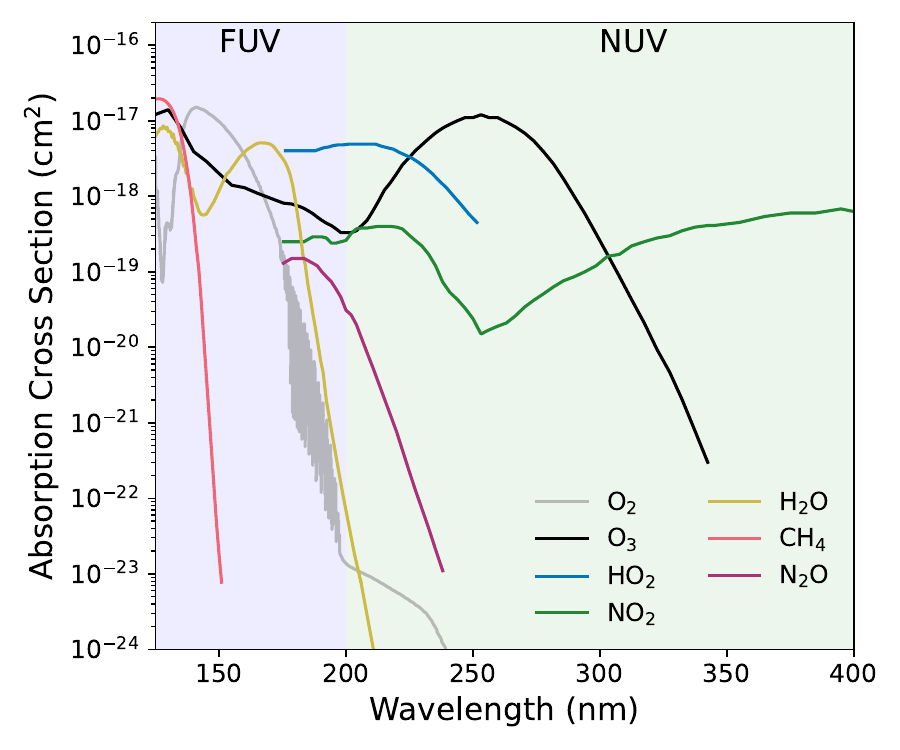}
    \caption{Stellar spectra for the host stars following \cite{koza22} (top) along with the cross-sections used by \texttt{Atmos} of relevant species (bottom). The ratio of far-UV (FUV) to near-UV (NUV) flux is important in driving atmospheric chemistry, and is indicated by the colored backgrounds. The abrupt cutoff of \noo, \no, and \hoo\ cross-sections is due to the extremely low photolysis rates of those species at shorter wavelengths due to absorption from other atmospheric species (i.e., CO$_2$). 
    \label{fig:stellar_spectra}}
\end{figure}

\section{Methods \label{sec:methods}}

\begin{table}[h!]
\centering
\caption{Model parameters. \label{tab:models}}
\begin{tabular}{lcc}
Model name & \ch\ MR    & \om\ MR \\
\hline
\hline 
\cite{koza22}           & 1.8$\times10^{-6}$ & 2.1$\times10^{-5}$ - 0.315 \\
\hline
Low \ch\ (10\% PAL)     & 1.8$\times10^{-7}$ & 2.1$\times10^{-5}$ - 0.315 \\
High \ch\ (1000\% PAL) & 1.8$\times10^{-5}$ & 2.1$\times10^{-5}$ - 0.315 \\
\hline
\hline
\end{tabular}
\tablefoot{
Abbreviations: MR = mixing ratio; PAL = present atmospheric level}
\end{table}

\subsection{Atmospheric models}
We used \texttt{Atmos}\footnote{https://github.com/VirtualPlanetaryLaboratory/atmos}, a publicly available 1D coupled photochemistry and climate code to model the atmospheres of Earth-like planets, following \cite{koza22} and \cite{koza25}. We briefly describe the code and our chosen parameters here, and refer the reader to other sources for a full description  \citep{arne16,mead18a,koza22}. Inputs to \texttt{Atmos} require a stellar host spectrum (121.6-45 450~nm), planetary parameters (e.g., radius, gravity), and the initial composition of the atmosphere and boundary conditions.

For the photochemistry code \citep{kast79,zahn06} we used the modern Earth template available with \texttt{Atmos}, including 50 gaseous species and 233 chemical reactions. The atmosphere is broken up into 200 plane parallel layers from the planetary surface to 100~km, with the flux and continuity equations solved simultaneously in each individual layer. Vertical transport is included for long-lived species, as well as molecular and eddy diffusion. Radiative transfer is calculated using a $\delta$-2-stream method \citep{toon89}, and the code is considered to be converged when the length of the adaptive time step reaches the age of the universe within the first 100 steps.

The climate code \citep{kast86,kopp13,arne16} then uses the atmospheric composition calculated in the photochemistry code along with the incoming stellar flux to calculate temperature and pressure profiles for the atmosphere. Here the atmosphere is broken up into 100 layers from the surface up until  1~mbar (typically $<$60-70 km), as the code does not run reliably at lower pressures \citep{arne16}. When transferring information back to the photochemistry code pressures above 1~mbar hold the temperature constant. Each layer uses a $\delta$-2-stream multiple scattering method to calculate stellar flux absorption. A correlated-$k$ method is used for outgoing IR for \oz, H$_2$O, CH$_4$, CO$_2$, and C$_2$H$_6$ with single and multiple scattering. Convergence is reached when the temperature and flux differences out of the top of the atmosphere are considered small enough ($\sim<$10$^{-5}$) \citep{arne16}.

For this study we iterated the photochemistry and climate codes back and forth 30 times and utilized the short-stepping technique for better climate code convergence \citep{teal22,koza22}. Our planets had the same radius and gravity as Earth, and orbited their host stars at the Earth-equivalent distance where they received the same total amount of flux as modern Earth. Fixed mixing ratio abundances of \om\ were varied from 0.01\% to 150\% PAL while additionally varying \ch\ (see Table~\ref{tab:models}).  We considered fixed \ch\ mixing ratios of 10\% and 1000\% PAL for high and low \ch\ scenarios (same as \no\ in \citealt{koza25}) to continue exploring the parameter space of possible Earth-like atmospheres and the resulting impact on the \om-\oz\ relationship. We note that for our high \ch\ models that the CH$_4$/CO$_2$ ratio is not high enough for haze formation \citep{arne16}, and it is therefore not considered in this study. Other relevant gaseous species held at constant mixing ratios of their present atmospheric levels are \no\ (3.0$\times10^{-6}$), H$_2$ (5.3$\times10^{-7}$), and CO (1.1$\times10^{-7}$).

\subsection{Input stellar spectra}
For host stars we used the same stellar spectra as in \cite{koza22} and \cite{koza25}, shown in Fig.~\ref{fig:stellar_spectra} along with absorption cross-sections of relevant species. For the G0V-K5V hosts the UV data came from \emph{International Ultraviolet Explorer} (IUE) data archives\footnote{http://archive.stsci.edu/iue} combined with synthetic ATLAS data for visible and IR wavelengths \citep{kuru79}. Our M5V host is GJ~876 from the Measurements of the Ultraviolet Spectral Characteristics of Low-mass Exoplanetary Systems (MUSCLES) survey \citep{fran16}. For full details of all host star spectra see \cite{rugh13} and \cite{koza22}.

\subsection{Post-processing radiative transfer models}
To explore the observational impacts of the changing \om-\oz\ relationship, we used the Planetary Intensity Code for Atmospheric Scattering Observations (\texttt{PICASO}) to simulate planetary emission spectra, using model atmosphere results from \texttt{Atmos} following \cite{koza22}. \texttt{PICASO}\footnote{https://natashabatalha.github.io/picaso/index.html}  \citep{bata19,bata21} is a publicly available code capable of simulating transmission, reflectance, and emission spectra, using atmospheric composition and temperature/pressure profiles calculated by atmospheric modeling codes. We simulated our model planets at full phase (0$^\circ$) from 0.3 to 14~$\mu$m, and focused in particular on the 9.6~$\mu$m \oz\ feature. We note that although it is unlikely for a planet to be imaged at full phase, it should not have a significant impact on mid-IR for a planet with minimal day-to-day temperature contrast.

\begin{figure*}[h!]
\centering
\includegraphics[scale=0.65]{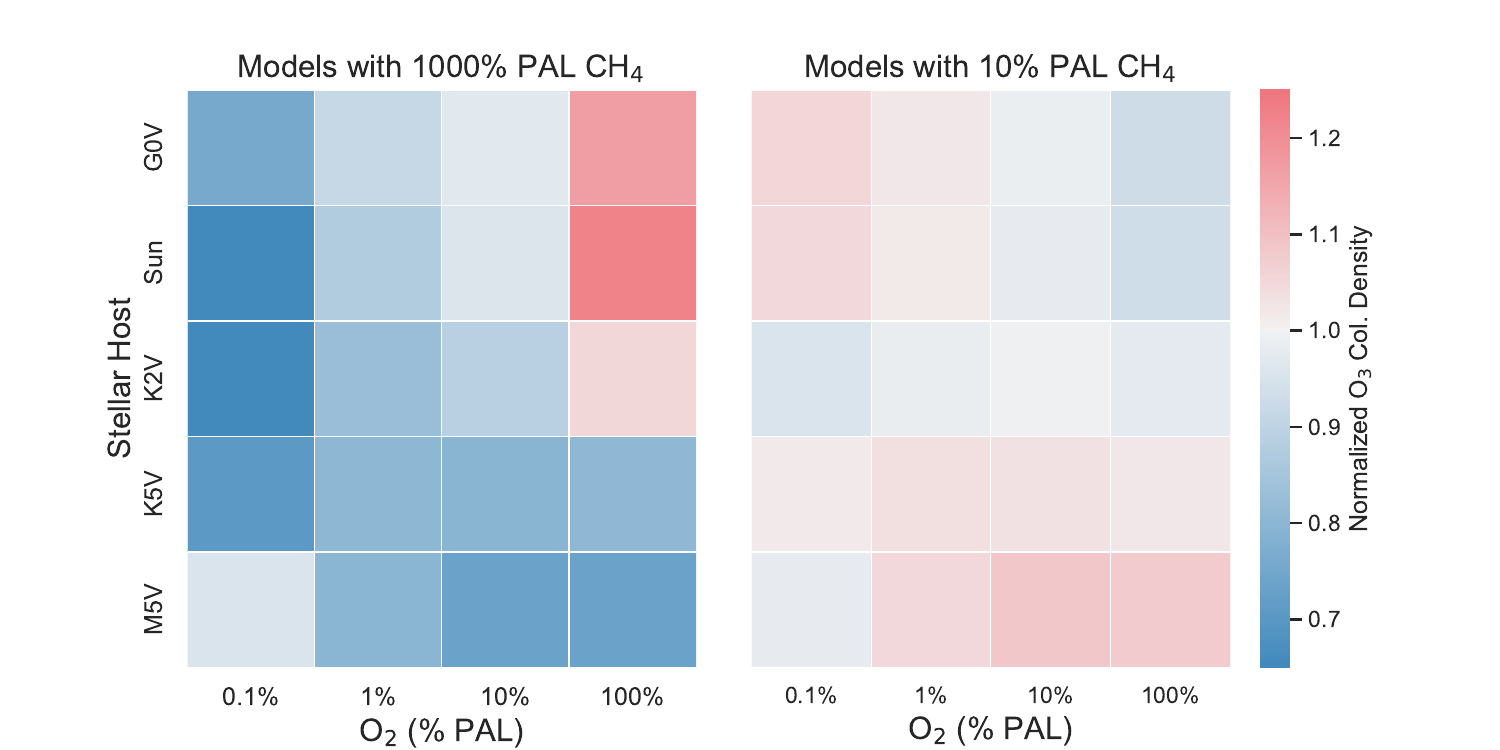}
   \caption{Ozone abundances normalized to the amount produced at modern levels of \ch\ for both the high (left) and low (right) \ch\ models for all host stars at 0.1\%, 1\%, 10\%, and 100\% PAL \om. The most striking result is that for the high \ch\ models there is less \oz\ compared to modern \ch, except for planets around the hottest hosts at 100\% PAL \om. This is due both to the indirect impact of \ch\ on the stratospheric temperature, as well as a boost in the smog mechanism due to a decreased \nox/\hox\ ratio. The rest of the cases for the \ch\ models show increased \oz\ depletion due to a higher amount of \ch\ from the increased efficiency of the \hox\ catalytic cycle's ability to destroy \oz.}
\label{fig:O2O3_relationship_grids}
\end{figure*}

\begin{figure*}[h!]
\centering
\includegraphics[scale=0.5]{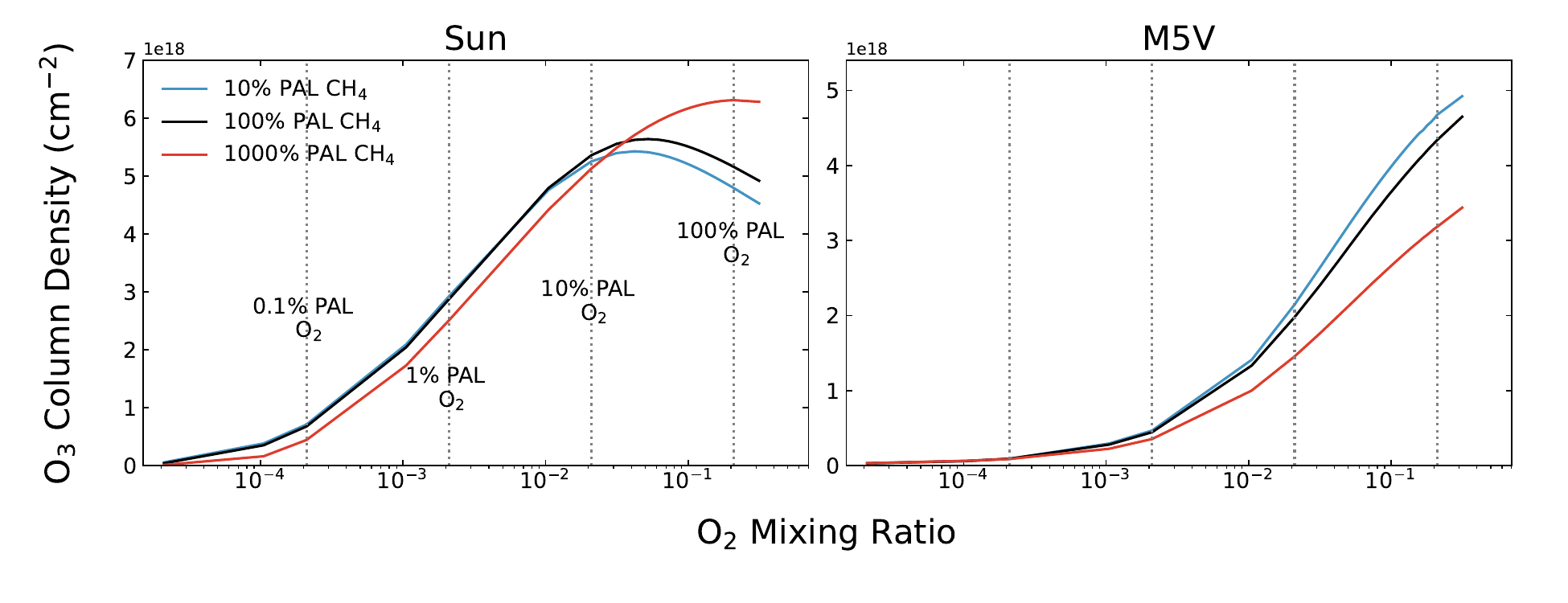}
\vspace{-0.5cm}
   \caption{Absolute values of the \om-\oz\ relationship at all \om\ and \ch\ levels modeled for planets around the Sun (left) and M5V (right) hosts. \om\ levels of 0.1\%, 1\%, 10\%, and 100\% PAL are marked with vertical dashed lines to enable easier comparison with Fig.~\ref{fig:O2O3_relationship_grids}. This figure highlights the stark difference in how \ch\ impacts the \om-\oz\ relationship differently for hotter and cooler host stars, due primarily to the amount of UV flux arriving at the planet. For hotter hosts the higher UV flux allows more efficient conversion of \ch\ into \hho\ and then \hox\ compared to the lower UV flux of M5V host. Full \om-\oz\ relationships of all hosts including comparisons to \cite{koza25} are located in Appendix~\ref{sec:appendix}.}
\label{fig:Sun_M5V_O2O3}
\end{figure*}

\section{Results \label{sec:results}}
We found that the impact of \ch\ on the \om-\oz\ relationship cannot be generalized, as it changes depending on the host star and the amount of \om. In Sect.~\ref{sec:CH4_chem} we explored the impact on atmospheric chemistry, changes in UV to the ground in Sect.~\ref{sec:UV_to_ground}, and impact on simulated planetary emission spectra in Sect.~\ref{sec:emission_spectra}. Supplementary figures and tables are available in Appendix~\ref{sec:appendix}.

\subsection{Atmospheric chemistry \label{sec:CH4_chem}}

\begin{figure*}[h!]
\centering
\includegraphics[scale=0.6]{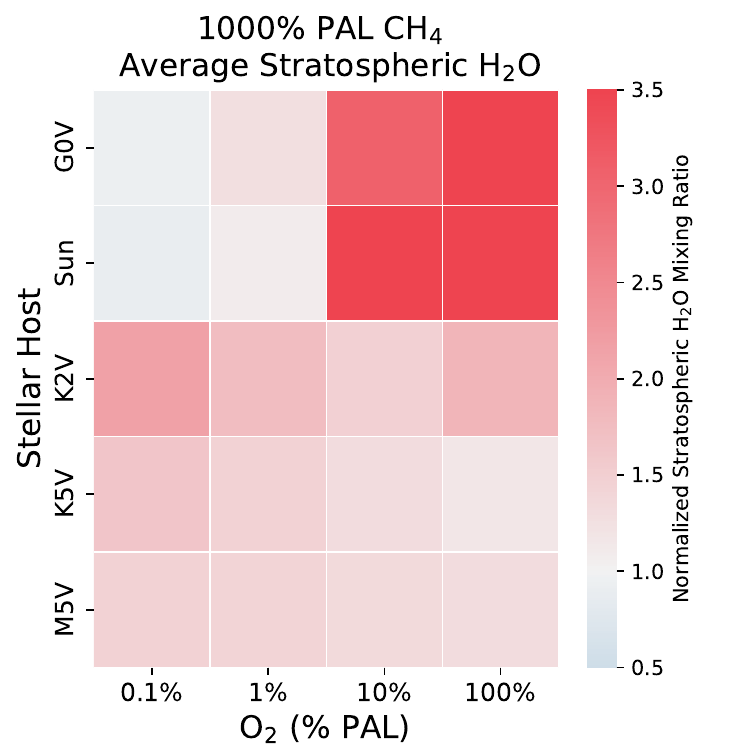}
\includegraphics[scale=0.6]{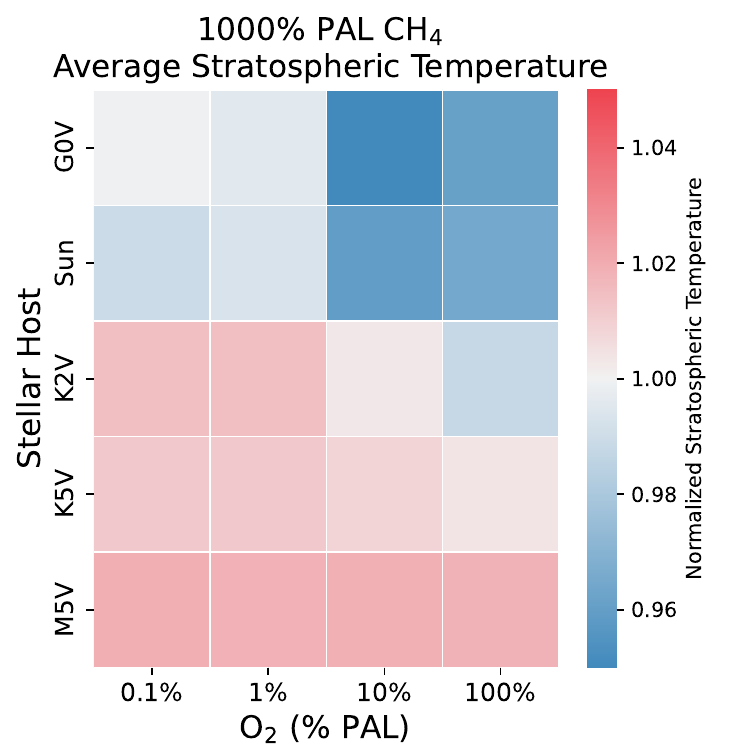}
\caption{Average stratospheric \hho\ (left) and temperatures (right) for high \ch\ models all normalized to results from models using modern levels of \ch. Nearly all models experience an increase in \hho\ for the high \ch\ models due to the conversion of \ch\ into \hho\ in the stratosphere. Planets around hotter hosts with higher \om\ show the largest increases as incident UV and \om\ are necessary for creating \hho\ in this scenario. The excess \hho\ impacts the temperature, providing stratospheric cooling in cases with large increases of \hho. As the atmosphere is thin in the stratosphere heat radiating from \hho\ can escape to space, causing an effect similar to stratospheric CO$_2$ cooling present on modern Earth. }
\label{fig:heatmap_H2O_temp}
\end{figure*}

\subsubsection{\emph{Atmospheric chemistry}: Overview}

Overall planet models were affected more by the high \ch\ models than the low \ch\ models, with higher \ch\ either depleting or increasing the total \oz\ abundance compared to atmospheres with modern levels of \ch\ depending on the host star and \om\ level, as seen in Fig.~\ref{fig:O2O3_relationship_grids} and Fig.~\ref{fig:Sun_M5V_O2O3}. The greatest increase in \oz\ from different \ch\ levels was experienced by the Sun-hosted planet with the high \ch\ model at 100\% \om\ PAL, resulting in 122\% of the original \oz\ abundance. The most \oz\ depletion also occurred with the high \ch\ models but for the K2V-hosted planet at 0.1\% PAL \om\ where it retains only 62\% of the original \oz\ abundance. 

An initially striking result is that for the high \ch\ models planets around all host stars at all \om\ levels display \oz\ depletion when compared to models with modern levels of \ch, except for those around the hottest host stars (G0V-K2V) at  \om\ levels similar to modern Earth. The increased \oz\ for the high \ch\ models around hotter hosts (as well as the decreased \oz\ for the corresponding cases with the low \ch\ models) might seem counterintuitive as \ch\ is the parent molecule for \hox, which catalytically destroys \oz\ (Sect.~\ref{sec:chemistryCH4}).  Indeed, \hox\ species are consistently more abundant with all high \ch\ models, powering more efficient \hox\ catalytic cycles that destroy \oz. The main processes affecting the \om-\oz\ relationship when varying \ch\ abundance are:
\begin{itemize}
    \item the amount of stratospheric \hho\ and \hox\ created by \ch,
    \item indirect effects of \ch\ on stratospheric temperature,
    \item smog mechanism efficiency as \nox/\hox\ ratio changes,
    \item changing \hox/\oz\ ratio at different \om\ levels.
\end{itemize}
We go into detail on each of these processes in the following subsections.

\subsubsection{\emph{Atmospheric chemistry}: Efficiency of converting \ch\ to \hox \label{sec:ch_to_hox}}
As summarized in Sect.~\ref{sec:chemistryCH4}, a \ch\ molecule can be converted into two H$_2$O molecules (Reactions~\ref{r:CH4_OH},\ref{r:CH3_O2},\ref{r:CH2O_OH}) given enough UV light and oxygen to power this process. In particular, OH, which begins this reaction chain, is formed by the \od\ radical reacting with H$_2$O (Reaction~\ref{r:H2O_O1D}), which is created via photolysis,
\begin{equation}
\m{O}_2 + \m{h}\nu \rightarrow \m{O }+ \m{O(}^1\m{D)  (}\lambda < 175\ \m{nm}),
\tag{\ref{r:PO2_O1D}}
\end{equation}
\vspace{-0.7cm}
\begin{equation}
\m{O}_3 + \m{h}\nu \rightarrow \m{O}_2 + \m{O(}^1\m{D)  (}\lambda < 310\ \m{nm}),
\tag{\ref{r:PO3_O1D}}
\end{equation}
\vspace{-0.7cm}
\be
\m{N}_2\m{O} + \m{h}\nu \rightarrow \m{N}_2 + \m{O(}^1\m{D) (}\lambda < 200 \m{ nm)},
\label{r:PN2O_O1D}
\ee
\vspace{-0.7cm}
\be
\m{CO}_2 + \m{h}\nu \rightarrow \m{CO} + \m{O(}^1\m{D) (}\lambda < 167 \m{ nm)},
\label{r:PCO2_O1D}
\ee
allowing planets with high amounts of incoming UV flux to be more efficient at converting \ch\ into \hho. In addition, the main source of \od\ radicals is typically from \oz\ photolysis (as it can be photolyzed at longer wavelengths than the other options), which causes the conversion of \ch\ into \hho\ to favor oxygen-rich environments. Due to the UV and oxygen requirements, conversion rates of \ch\ into \hho\ and then into \hox\ are faster for planets around hosts with higher UV flux at higher \om\ levels. 

The higher levels of \hox\ that the high \ch\ models bring cause faster \hox\ catalytic cycles that destroy \oz. However, planets around the 3 hottest hosts at 100\% PAL \om\ experience an increase in \oz\ for the high \ch\ models, despite the fact that they are the hosts converting the most \ch\ into \hox. The reason behind this \oz\ increase is due to the indirect impact of \ch\ on stratospheric temperature, and the efficiency of the smog mechanism for planets around those hosts with high amounts of \hox.

\subsubsection{\emph{Atmospheric chemistry}: Indirect impact of \ch\ on stratospheric temperature}
Planets around hotter hosts at high \om\ levels are the most efficient at converting \ch\ into \hox, but also into \hho. Although this increases the amount of \hho\ converted into \hox, another effect of the large amount of stratospheric \hho\ is that it forms at a high enough altitude to have a cooling effect. Unlike in the troposphere, where \hho\ heats the atmosphere as a greenhouse gas, in the stratosphere \hho\ radiates heat into space, similar to CO$_2$ stratospheric cooling seen on modern Earth \citep{goes16,sant23}. In the stratosphere, \hho\ radiates more in the infrared than absorbs energy coming from the lower atmosphere, resulting in a net cooling. This causes atmospheres with enough excess \hho\ to experience significant stratospheric cooling, especially for planets with hotter hosts at high \om\ as shown in Fig.~\ref{fig:heatmap_H2O_temp}. For the low \ch\ cases the opposite occurs where there is less stratospheric \hho, causing a warmer stratosphere when compared to modern levels of \ch, although the effect is smaller than for the high \ch\ cases.

Stratospheric cooling caused by excess \hho\ for the hottest hosts at high \om\ causes two main changes in the \oz\ abundance:
\begin{itemize}
    \item  a faster Chapman mechanism and faster \oz\ production,
    \item a slower \nox\ catalytic cycle and slower \oz\ destruction.
\end{itemize}
Stratospheric cooling from increased \hho\ abundance occurs for a range of \om\ levels for the planets around the G0V and Sun hosts, but only to the extent that it overcomes the depletion of \oz\ from the faster \hox\ catalytic cycles at high \oz\ with high \ch.

\subsubsection{\emph{Atmospheric chemistry}: Smog mechanism efficiency with changing \hox \label{sec:smog}}
In addition to the faster \oz\ production and slower \oz\ destruction due to stratospheric cooling from excess \hho\ with the high \ch\ models, planets hosted by the hottest stars also experience a boost in \oz\ production from the smog mechanism. Although \hox\ is not directly created or consumed by the smog mechanism (Reactions~\ref{r:O2M},\ref{r:OH_CO},\ref{r:H_O2},\ref{r:HO2_NO},\ref{r:PNO2}), it is necessary to the process as a catalyst. As discussed in both Sects~\ref{sec:oz_formation} and~\ref{sec:NOx_regimes}, at high enough levels of \nox\ the atmosphere will enter the \nox-saturated regime and \nox\ will begin to deplete \hox\ by converting it into reservoir species (see Fig.~\ref{fig:NOx_regimes}). Planets around hotter stars (G0V-K2V) in particular have enough \nox\ in their lower stratospheres to place them in the \nox-saturated regime for modern levels of \no, resulting in significant \hox\ depletion.

\begin{figure}[h!]
\centering
\includegraphics[scale=0.3]{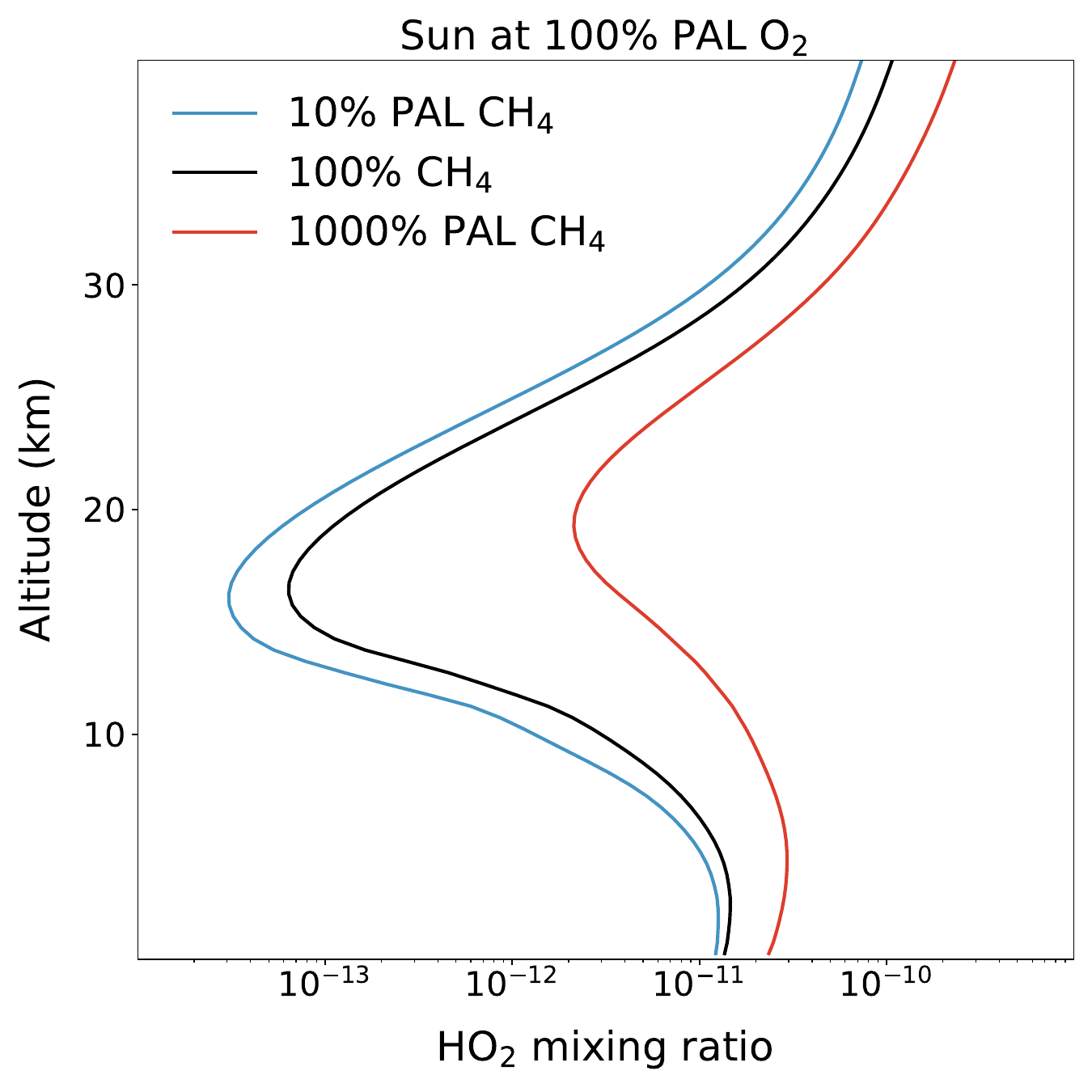}
\caption{Profiles of \hoo\  for the planet around the Sun at 100\% PAL \om\ and different levels of \ch. The significant \hoo\ depletion present for low and modern levels of \ch\ is lessened for the high \ch\ models. This is because the \hoo\ depletion is due to the Sun-hosted planet being in the \nox-saturated regime, but increased \ch\ allows enough \hox\ production to lessen this effect.
\label{fig:Sun_CH4_HO2profiles}}
\end{figure}

In \cite{koza25} we determined which \nox\ regime an atmosphere was in by looking at the \nox\ abundance, but in this study we took a different approach because it is \hox, not \nox, that has significant variation. In environments where there are greatly increased levels of \hox, we find that the cutoff of \nox\ that determines the \nox-saturated and \nox-limited regimes changes. It is seen in Fig.~\ref{fig:Sun_CH4_HO2profiles} for the planet around the Sun at 100\% PAL \om\ with the high \ch\ models that there is a significant decrease in depletion of \hoo\ caused by high \nox\ when compared to models with modern levels of \ch. Although the planet around the Sun at modern levels of \om\ and \nox\ was squarely in the \nox-saturated regime, since the \nox/\hox\ ratio is smaller with the increased \ch\ there is significantly less \hoo\ depletion indicating that the atmosphere has been pushed toward the \nox-limited regime. We see evidence of this occurring for the G0V, Sun, and K2V hosts at \om\ levels near 100\% PAL \om\ with the high \ch\ models, resulting in a boost of smog mechanism-produced \oz. This is the other part of the puzzle as to why planets around these hosts experience an increase in \oz\ with the high \ch\ models at modern \om\ levels.

The smog mechanism is also the reason that the M5V-hosted planet with the high \ch\ models experiences less \oz\ depletion as \om\ levels decrease, as opposed to planets around all other hosts which experience more \oz\ depletion with decreasing \om. The planet around the M5V host is the only one to always exist in the \nox-limited regime, and therefore does not suffer from suppression of the smog mechanism as do planets around the other hosts. In addition, smog formation becomes even more efficient at lower \om\ levels as \hox\ is ``pushed'' closer to the ground as UV shielding from \om\ and \oz\ decreases. Photolysis allows \hox\ to form deeper in the atmosphere, allowing the lower atmosphere \hox\ to speed up smog formation even further (see \citealt{koza22} for a longer discussion of this process).  This effect occurs for planets around all host stars, but unlike planets around other hosts, for the M5V-hosted planet the extra \oz\ produced from the smog mechanism has a more significant impact on the \om-\oz\ relationship since smog-produced \oz\ makes up a much larger portion of the total \oz\ abundance. This is because the Chapman mechanism has a photon requirement of less than 242~nm for \om\ photolysis, whereas the smog mechanism can be fulfilled with near-UV and visible photons for the required \noo\ photolysis (see Figure~\ref{fig:stellar_spectra}). The low UV flux of the M5V host allows the smog mechanism to take on a larger role in \oz\ production than for planets around hotter hosts. For this reason for the high \ch\ models with the M5V-hosted planet the amount of \oz\ depletion decreases with decreasing \om. As the amount of Chapman-produced \oz\ from \om\ photolysis decreases, increased amounts of smog-produced \oz\ from added \hox\ are more significant.

\subsubsection{\emph{Atmospheric chemistry}: Changing \hox/\oz\ ratio}

For planets around all hosts other than the M5V as the \om\ level decreases, a larger portion of \oz\ is depleted. This is because as \om\ abundance decreases, the \hox/\oz\ ratio increases. As discussed in Sect.~\ref{sec:ch_to_hox}, conversion from \ch\ into \hox\ is more efficient at higher \om\ levels as \oz\ is often an indirect source of \hox. Formation of \hox\ occurs when \hho\ reacts with \od\ radicals, which are most easily formed from \oz\ photolysis when there is sufficient \oz\ supply, as \oz\ can be photolyzed by much longer wavelength photons than the other suppliers of \od: \om, \no, and CO$_2$ (Reactions~\ref{r:PO2_O1D},\ref{r:PO3_O1D},\ref{r:PN2O_O1D},\ref{r:PCO2_O1D}). However, when \oz\ (and \om) levels decrease, \hox\ begins to source \od\ radicals more from photolysis of \no\ and CO$_2$. Although this requires higher energy photons, it is also easier for these species to be photolyzed deeper into the atmosphere as \om\ and \oz\ levels decrease (causing less UV shielding), thus creating a larger \hox/\oz\ ratio with decreasing \om. With a higher \hox/\oz\ ratio the \hox\ catalytic cycle has an easier time depleting \oz\ at these lower \om\ levels. As discussed in the previous section (Sect.~\ref{sec:smog}) this effect of increasing amounts of \oz\ depletion occurring at lower \om\ levels does not occur for the M5V-hosted planet due to the boost in smog-produced \oz.

\begin{figure*}
\centering
\includegraphics[scale=0.45]{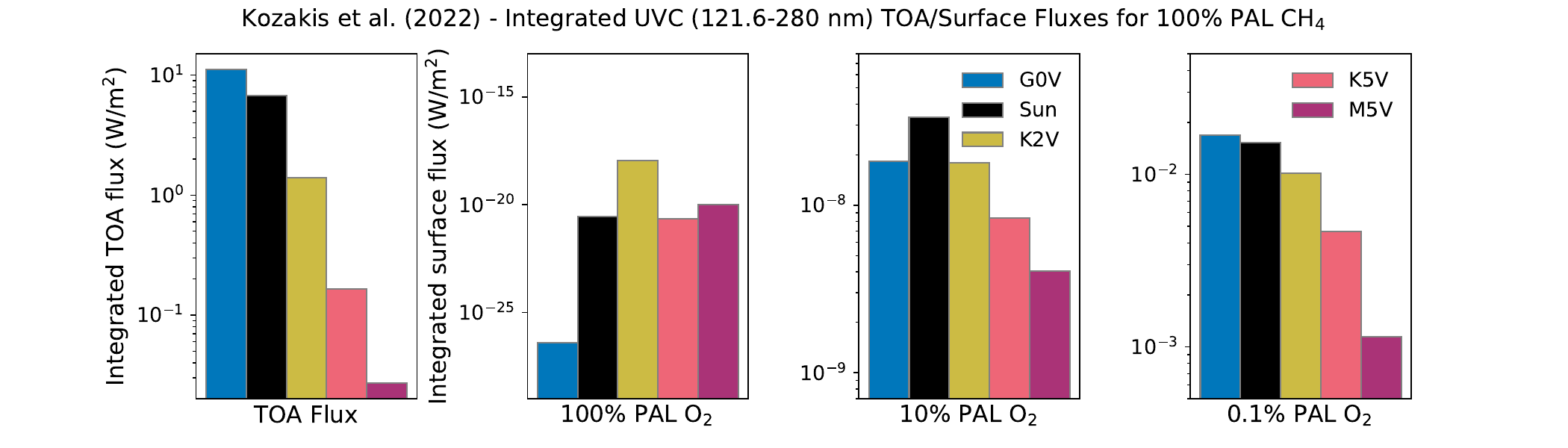}\\
\includegraphics[scale=0.45]{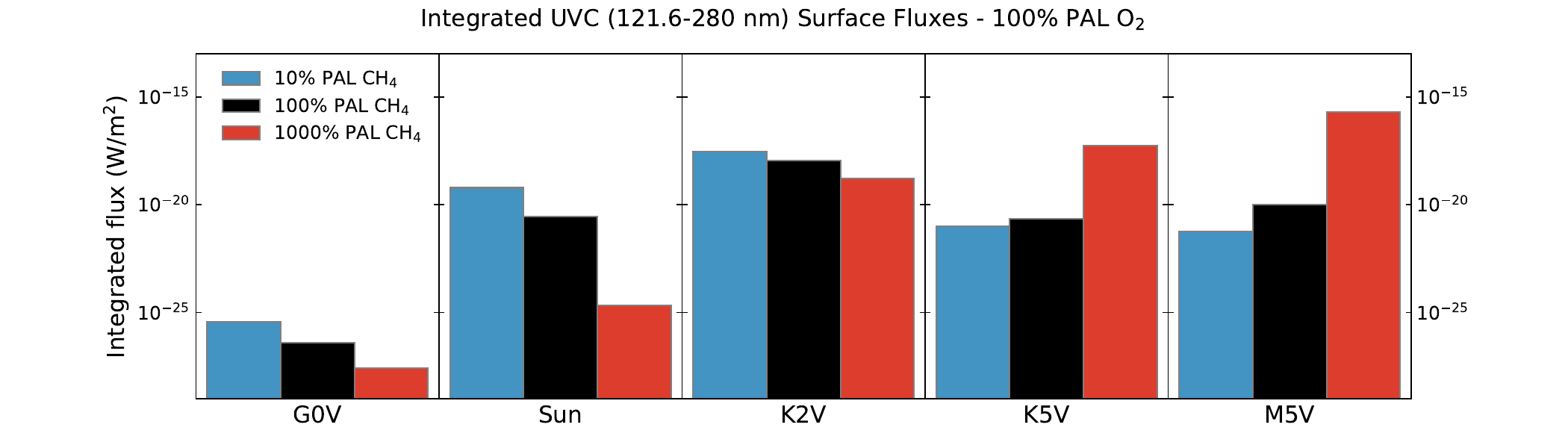}\\
\includegraphics[scale=0.45]{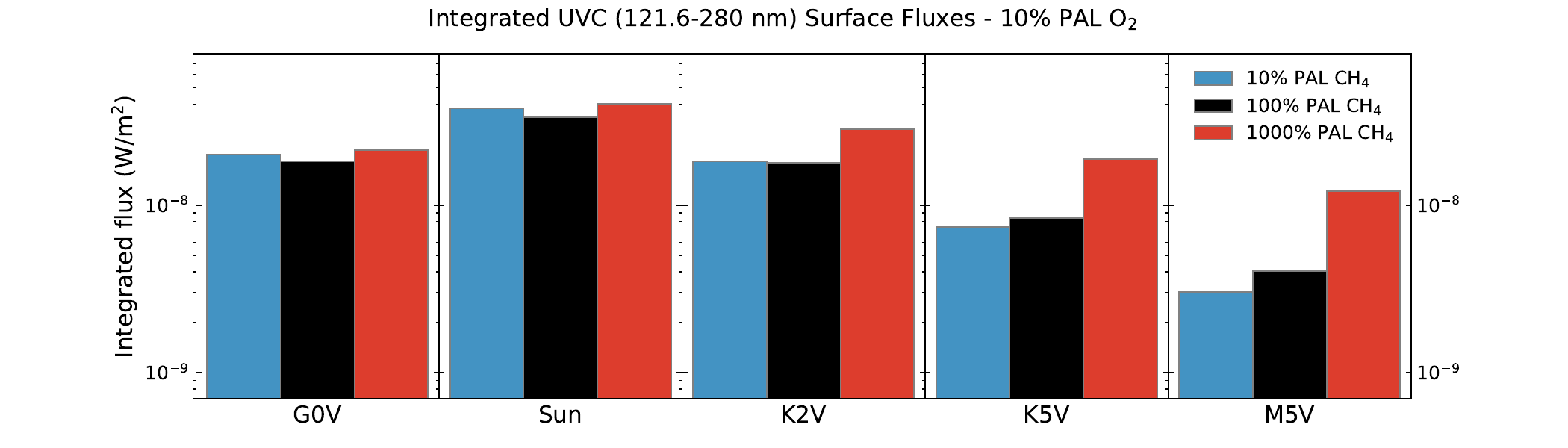}\\
\includegraphics[scale=0.45]{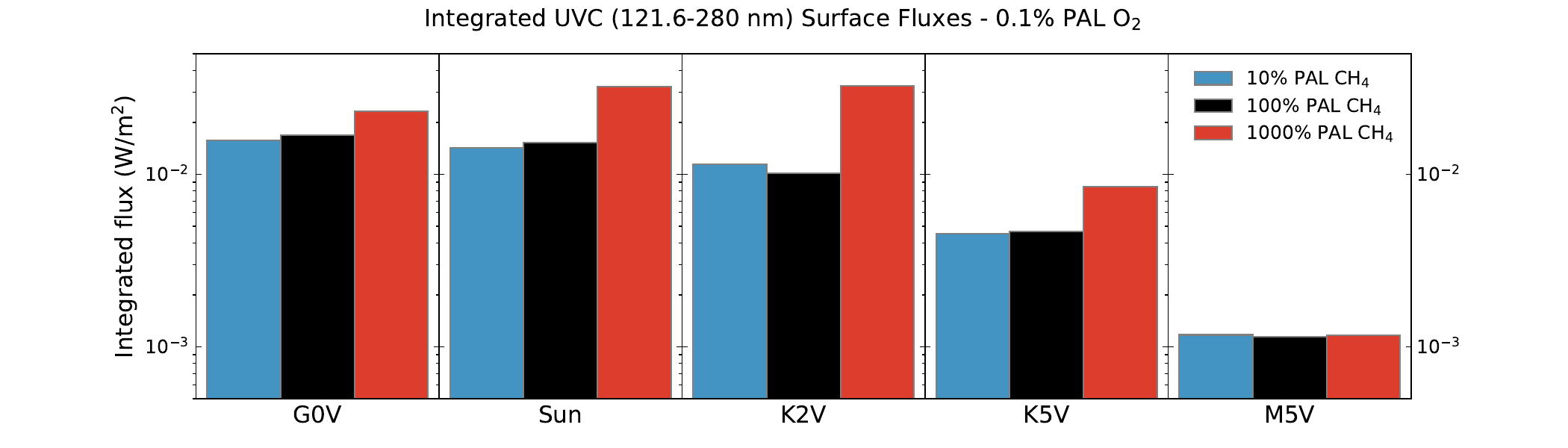}\\
    \caption{Comparisons of UVC results with modern \ch\ abundances (top row) with incident top-of-atmosphere (TOA) UVC flux and surface UVC flux for all hosts with modern levels of \ch\ at different \om\ levels, and surface UVC flux with varying \ch\ for 100\% PAL \om\ (second row), 10\% PAL \om\ (third row) and 0.1\% PAL \om\ (bottom row). Surface UVC plots from \cite{koza22} use the same y-axis limits as the corresponding plots in the bottom three rows to enable easier comparison. Changes in UVC surface flux are most significant for 100\% PAL \om\ due to the dependency of oxygen to convert \ch\ into \hho\ and \hoo.}
\label{fig:UVC_barchart_CH4}
\end{figure*}

\subsection{UV to ground \label{sec:UV_to_ground}}

Variations in \oz\ caused by different \ch\ abundances resulted in varying amounts of UV shielding in the atmosphere, and therefore different amounts of UV flux reaching the surface of our model planets. The level of potential biological damage that UV photons can inflict is dependent on wavelength, described by three UV regimes that we used here. UVA (315-400~nm) is the lowest energy UV and least dangerous, and is not shielded by \oz; UVB (280-315~nm) is responsible for skin tanning and skin cancer, and is partially shielded by \oz; and lastly UVC (121.6-280~nm) is capable of significant biological harm, but is fortunately well shielded by \oz\ and \om, assuming that they exist in significant quantities in the atmosphere. Along with being the most dangerous, UVC surface flux also has a highly nonlinear relationship with the amount of \oz\ in the atmosphere, as it is UVC flux ($<$~242~nm; Reactions~\ref{r:PO2_O}, \ref{r:PO2_O1D}) that is necessary for the Chapman mechanism to produce \oz. Results for UVC surface flux variations at 100\%, 10\%, 1\%, and 0.1\% PAL \om\ are shown in Fig.~\ref{fig:UVC_barchart_CH4}, along with a table for UVB and UVC results in the Appendix (Table~\ref{tab:UV_all}). As in \cite{koza22} UVA surface flux did not vary with changing \om\ or \ch\ values (always $\sim$80\% reaching the surface) so it is not discussed here.

As UVB flux is partially shielded by \oz, varying \ch\ causes changes to the amount of UVB photons reaching the ground in some cases, but only very slightly. The largest decrease was  for the Sun-hosted planet at 100\% PAL \om\ with the high \ch\ models having only 80\% of the original UVB surface flux due to increased \oz\ shielding. The largest increase in UVB surface flux was also at 100\% PAL \om\ for the high \ch\ models, but this time with the M5V-hosted planet receiving 124\% the UVB surface flux that it did with modern levels of \ch. For \om\ levels under 100\% PAL changes in the amount of UVB flux arriving at the surface changed only slightly, with larger effects being seen consistently with the high \ch\ models due to the larger impact on \oz\ formation and destruction (see Table~\ref{tab:UV_all}).

Changes in UVC surface flux were much more significant, due to the larger absorption cross-sections at these wavelengths. The highest increases and decreases of UVC surface flux corresponded to the same models with the largest UVB surface flux differences: the planets around the Sun and M5V hosts, both for the high \ch\ models at 100\% PAL \om. The planet around the Sun  experienced a factor of just 7.5$\times10^{-5}$ times the original UVC flux at modern levels of \ch\ due to the increased amounts of \oz\ caused by stratospheric cooling, decreased \nox\ catalytic cycle efficiency, and extra smog production. The largest increase in UVC surface flux from the planet around the M5V host (receiving a factor of 2.1$\times10^{4}$ more) was due to the increased ability for the \hox\ catalytic cycle to destroy \oz. For the low \ch\ models the largest variations in UVC surface flux were similarly at 100\% PAL \om\ with the largest increase in surface UVC for the planet around the Sun with a factor of 22 times more UVC surface flux, and the largest decrease was for the M5V-hosted planet having 5.8$\times10^{-2}$ times the original UVC flux due to the decreased ability of the \hox\ catalytic cycle to destroy \oz. For lower oxygen levels the variation in \oz\ was significantly decreased, causing the variation in UVC surface flux to be minimized, with the amount of UVC flux variation compared to modern \ch\ levels not exceeding an order of magnitude.

\begin{figure*}
\centering
\includegraphics[scale=0.5]{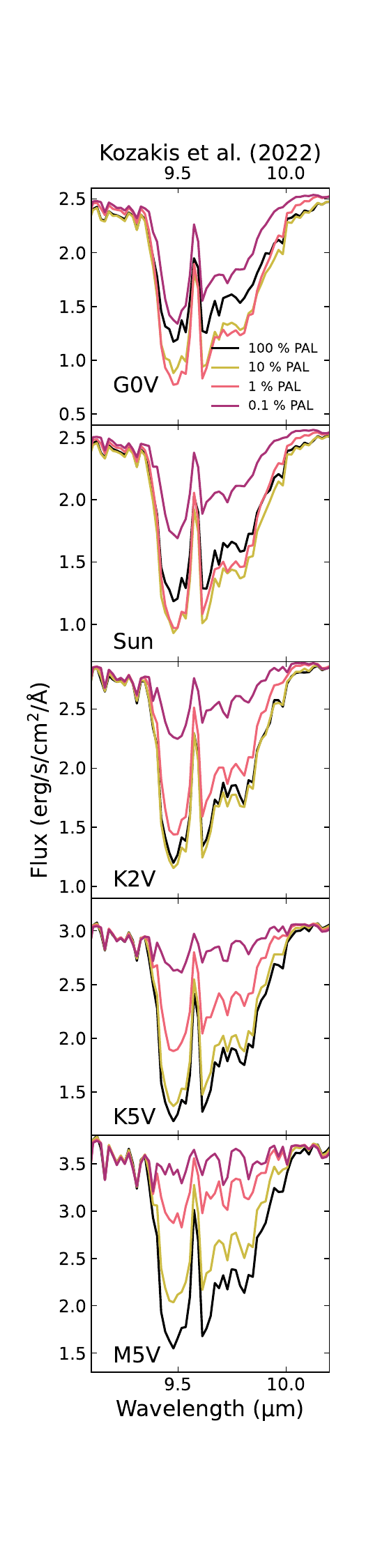}
\hspace{-1.2cm}
\includegraphics[scale=0.5]{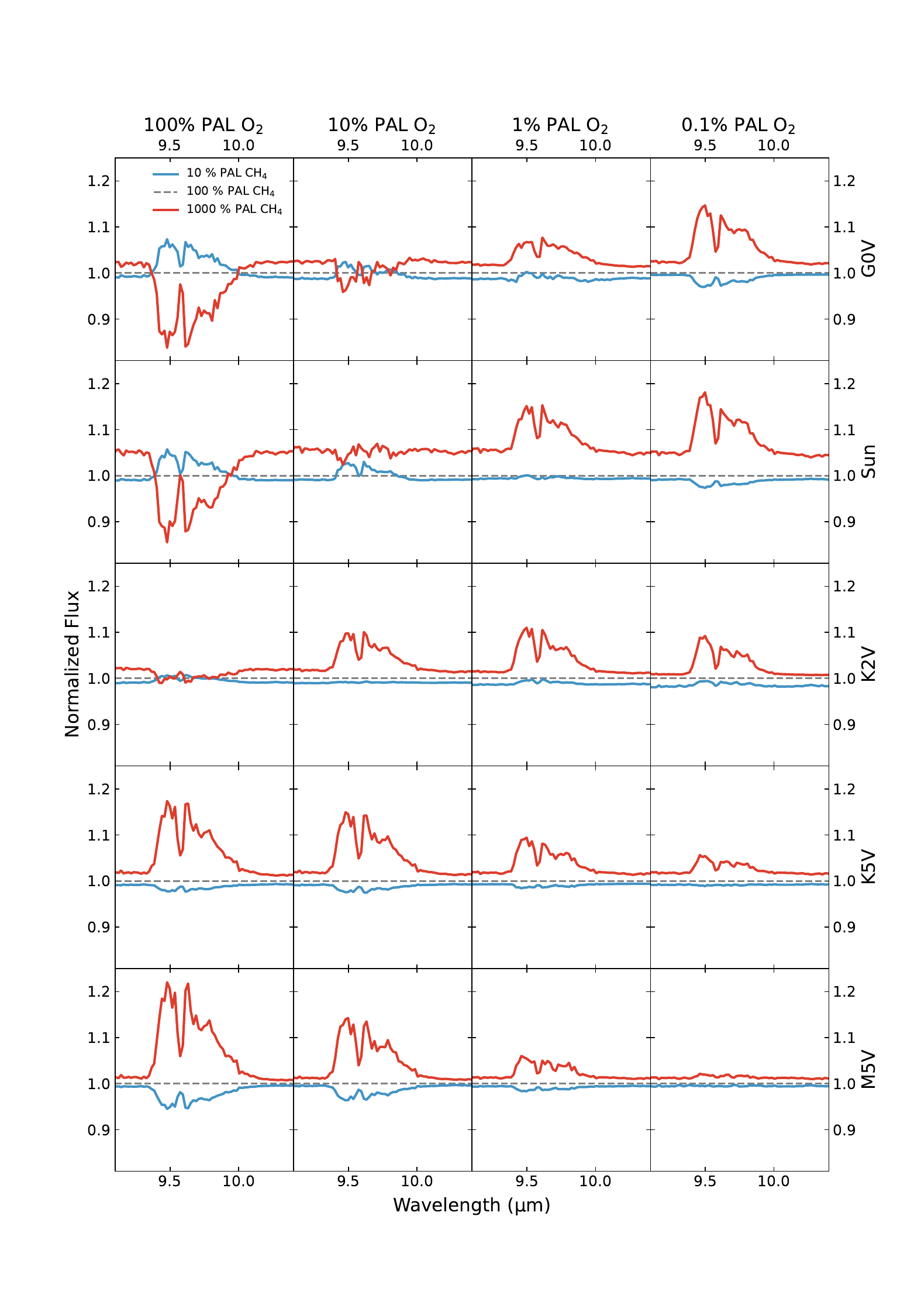}
\vspace{-1.5cm}
\caption{Comparisons of the 9.6~$\mu$m \oz\ emission spectra feature for all hosts at different \om\ levels using modern levels of \ch\ (left) and \oz\ features for different \ch\ abundances (right) normalized to features with modern levels of \ch.  Normalized \oz\ features use the same y-axis limits for all host stars in order to compare the difference in impact of \ch\ on \oz\ for different hosts. For the hottest hosts at 100\% PAL \om\ changes in feature strength are due to changes in \oz\ abundances as well as stratospheric temperatures differences from \hho\ abundance, while the rest of the model spectra variations are due mainly to changes in \oz\ due to variations in \ch.
}
\label{fig:emission}
\end{figure*}

\subsection{Planetary emission spectra \label{sec:emission_spectra}}
We additionally generated planetary emission spectra to explore the potential impact on future observations that varying \ch\ could have on the 9.6~$\mu$m \oz\ feature. Emission spectral features are highly influenced by the temperature difference between the emitting and absorbing layers of the atmosphere, which as shown in \cite{koza22} is particularly relevant when considering \oz\ features. Since \oz\ NUV absorption is the primary source of stratospheric heating on modern Earth, there is a counterintuitive effect where a planet with large amounts of \oz\ can have a shallower emission feature than a planet with significantly less \oz. This is because a planet experiencing stratospheric heating via \oz\ will have a decreased temperature difference between the planetary surface and stratosphere, resulting in a shallower emission spectral feature depth. However, this will only happen for planets receiving sufficient NUV flux from their host, as this is required for said stratospheric heating to occur. This is shown in the left panel of Fig.~\ref{fig:emission}, where variations in the 9.6~$\mu$m \oz\ feature from just varying \om\ in \cite{koza22} are shown. For a more detailed discussion of this phenomena, see \cite{koza22}.

Once again it the high \ch\ models had more of an impact than the low \ch\ models (see Fig.~\ref{fig:emission}). Planets around hotter hosts experienced the most variability in the \oz\ feature due to different \ch\ and \om\ abundances, owing to the indirect impact of \ch\ on stratospheric temperature.  For planets that experienced stratospheric cooling from high \hho\ content, \oz\ spectral features are much deeper both because of increased \oz\ abundance, but also from the larger temperature difference between the emitting and absorbing layers of the atmosphere. However, at lower \om\ levels for planets around these same hosts the \oz\ features became shallower than with modern \ch\ abundances, due to lower \oz\ abundances from the higher destruction rates of \oz\ via more productive \hox\ catalytic cycles. For all other cases the high \ch\ models resulted in shallower \oz\ spectral features, again due to the decreased amounts of \oz\ from the higher efficiency of the \hox\ catalytic cycle. Stratospheric temperatures did not vary significantly in these cases, leading to changes in the \oz\ feature depth corresponding more directly to changes in \oz\ abundance. It is also worth noting that atmospheres with the high \ch\ models had generally hotter surface temperatures and atmospheres as \ch\ is a greenhouse gas. Overall feature depth changed the most for planets around the G0V and Sun hosts with the high \ch\ models, as they experienced significant deepening of features at high \om\ levels, and much shallower features at low \om\ levels.

\section{Discussion \label{sec:discussions}}

\subsection{Comparison to impact of \no\ on the \om-\oz\ relationship}
This paper is in many ways a ``sister study'' to \cite{koza25}, which varied \no\ instead of \ch\ in order to understand how it would impact the \om-\oz\ relationship. Both \no\ and \ch\ are potential biosignatures \citep{schw18,schw22,thom22,ange24}, power the primary catalytic cycles (with \nox\ and \hox), and influence the smog mechanism of \oz\ formation. However, we found that Earth-like atmospheres experience different responses to variations and \no\ and \ch, depending on both the amount of \om. As the ultimate goal of this paper series is to explore and navigate the challenges of using \oz\ as a proxy for \om, it is useful to compare how \no\ and \ch\ impact the \om-\oz\ relationship differently. Figures comparing results from \cite{koza25} and this study are located in Appendix~\ref{sec:appendix}.

Overall at higher \om\ levels ($>\sim$1\% PAL \om) varying \no\ had a stronger impact on the total amount of \oz\ for planets around all stars except for the M5V host, due to changing efficiency of the \nox\ catalytic cycle (Fig.~\ref{fig:O2O3_relationship}). For the high \no\ models at high \om\ the depletion of \oz\ was the most extreme, causing orders of magnitude differences in harmful UVC flux reaching the ground -- significantly larger changes than when varying \ch\ (Fig.~\ref{fig:UVbarchart} and Table~\ref{tab:UV_all}). For lower \om\ levels high \ch\ led to \oz\ depletion, while high \no\ resulted in increased \oz\ formation from a boost to the smog mechanism. Changes in the \oz\ abundance at these lower \om\ levels from the low \no\ and \ch\ models were much smaller. Both \no\ and \ch\ were found to impact the smog mechanism, demonstrating in different scenarios the importance of the \nox/\hox\ ratio in determining if smog production would be enhanced or suppressed. However, it was only variations in \ch\ that induced atmospheric changes that could significantly alter atmospheric temperature profiles (through production or lack of production of stratospheric \hho).

Looking at the 9.6~$\mu$m \oz\ feature from the simulated emission spectra (Fig.~\ref{fig:emission_N2O_CH4}), it is clear that planets around every host star were impacted from the variations in either \no\ either/or \ch. The cases least affected were around the coolest hosts at the lowest \om\ values. The most significant changes overall in the \oz\ spectral feature were for the K2V-hosted planet at 100\% PAL \om\ with high \no, which was ironically the only case that was relatively untouched by \ch\ variations at 100\% PAL \om. Planets around the G0V and Sun hosts faced the most variations in how \no\ and \ch\ impacted the \oz\ feature based on the \om\ level, with high \no\ causing shallower features at high \om\ and deeper features at low \om, with the reverse occurring for the high \ch\ models. We also find that the \oz\ feature of the M5V-hosted planet was only significantly changed by variations in \ch, not in \no. Planets around all other hosts experienced spectral feature changes from both \no\ and \ch\ variations to some degree. For future observations it seems that understanding the \no\ and \ch\ content of a planetary atmosphere will be important if we wish to use \oz\ as a proxy for \om, although our results thus far indicate that changes in \no\ will be less relevant for planets around hosts like the M5V star used here, but that \ch\ can have much more of an effect on \oz\ abundance.

\subsection{Comparisons to other studies}
Although this is the first study to vary both \om\ and \ch\ in Earth-like atmospheres orbiting a variety of host stars with the goal of studying the impact on \oz, there are several other studies exploring similar concepts. \cite{gren06} varies both \ch\ and \om\ abundances (along with \nox, H$_2$, and CO) in the context of Proterozoic Earth to evaluate the role of smog-produced \oz\ in UV shielding during that time period. They focus on \om\ levels used in this study (10\% and 100\% PAL \om) but adopt \ch\ mixing ratios an order of magnitude larger than we used with our high \ch\ models (1.0$\times10^{-4}$ and 3.0$\times10^{-4}$). Therefore we cannot compare directly to our results for the planet around the Sun, but we do see some similar trends. \cite{gren06} found that \nox\ was the main driver in changing \oz\ production, but that when \nox\ abundances were high enough to enter the \nox-saturated regime, increasing \ch\ (and therefore \hox) provided a boost in the smog mechanism. This agrees with our results for the hottest hosts in our study at high \om\ (and therefore the highest \nox), as we found that increased \ch\ and \hox\ helped to counteract \nox-saturated atmospheres and increased the amount of smog-produced \oz\ (see Fig.~\ref{fig:Sun_CH4_HO2profiles}). Despite the different parameter spaces in \cite{gren06} and this study, both display the same trends of the importance of the \nox/\hox\ ratio determining the efficiency of the smog mechanism, and that in high \nox\ environments extra \hox\  results in more smog-produced \oz.

\cite{gren14} also varies \ch\ on an Earth-like planet, but around an M7V host star. They focus on the impact on biosignatures when varying biological surface fluxes of \no\ and \ch, along with the UV spectrum of the host star, using \ch\ fluxes 100 times less than present day, and 2 and 3 times that of present day. Although they see some similar trends to the work in this study (and results agreeing well with \citealt{koza25}) the differences in \ch\ boundary conditions (fixed surface flux rather than fixed mixing ratio) and different UV abundances make it difficult to draw many direct comparisons. That being said, there are no direct contradictions to the findings of this study, with differences in results likely due to different parameter space and boundary conditions.

Searching through the literature we could not find any studies replicating our results of stratospheric cooling via excess \hho\ with our high \ch\ models around hotter hosts, although several studies find high amounts of \hho\ production in high \ch\ atmospheres (e.g., \citealt{segu05,raue11}). We do not believe our results of stratospheric cooling contradict any existing studies as this is the first to model such levels of \om\ and \ch\ around hotter hosts, and stratospheric cooling due to greenhouse gases such as CO$_2$ has been modeled and observed on modern Earth (e.g., \citealt{goes16,sant23}).

We also note that the decision to use fixed mixing ratios as the boundary condition for certain species (\om, \no, \ch, H$_2$, CO) impacts our results differently than if we were to use fixed surface fluxes. As discussed in \cite{koza22} and \cite{koza25}, species such as \no\ and \ch\ have been shown to build up in the atmospheres of Earth-like planets around cooler host stars due to the low incident UV flux (e.g., \citealt{rugh15,wund19,teal22}). In addition, CO abundance (which is strongly interlinked with \ch\ abundance) has been shown to similarly build up in such environments (see \citealt{schw19} and references therein). We reiterate here that our choice in fixed mixing ratio boundary conditions is to enable easier comparison between the differences in the \om-\oz\ relationship for planets around different host stars. The impact of different boundary conditions will be explored at length in future work.

\subsection{Plausible \ch\ mixing ratios in Earth-like atmospheres}
The purpose of this study was to further evaluate potential changes in the \om-\oz\ relationship when \ch\ levels are varied, and our choice of using \ch\ abundances of 10\% and 1000\% PAL was motivated by the need to begin filling out the parameter space over which Earth-like exoplanets may exist. The plausible range of \ch\ in the atmospheres of such planets is still unknown, but we can use our knowledge of \ch\ abundances over the geological history of Earth as a rough guide.

A complication to making such predictions is that surface flux and resulting atmospheric mixing ratios of \ch\ are nonlinear (similar to \no), with a strong dependency on the abundance of \om\ and other oxidizing species. \om\ combined with \ch\ is said to be a promising disequilibrium biosignature pair due to the fact that they react quickly with each other and large surface fluxes of \om\ and \ch\ are required to allow significant amounts of both to co-exist in an Earth-like atmosphere. The amount of \ch\ that is able to accumulate in the atmosphere is additionally dependent on the type of host star, where similarly to \no, planets around cooler hosts with less incoming UV allow for a larger buildup of \ch\ (e.g., \citealt{segu03,rugh13}). 

On the early Earth before the Great Oxidation Event \ch\ was likely significantly higher in abundance due to the reducing atmosphere and lack of appreciable \om, and it is possible that during the Archean era that \ch\ abundances could have been greater than 1000 ppm (e.g., \citealt{arne16,catl20}). \cite{thom22} and \cite{akah24} both modeled surface fluxes of \ch\ to see the corresponding \ch\ atmospheric mixing ratios around the Archean Earth, before the rise of \om, around the Sun and other spectral hosts. They found that during this time period although there was very little \om, \ch\ still faced destruction from photolysis and reactions with OH created from H$_2$O photolysis.

After the rise of \om\ it is possible that there was still a significant amount of \ch\ in the atmosphere, perhaps contributing significantly to the warming of the Proterozoic Earth \citep{robe11}, although it also is possible that there was not significant \ch\ during this era (e.g., \citealt{olso16}). The relationship between \ch\ surface flux and atmospheric mixing ratios has been explored at length in \cite{greg21}, which varied both \ch\ and \om\ surface fluxes, as they highly impact the stability of both species in the atmosphere. In \cite{koza25} we were able to discuss an actual limit of biologically produced \no, but with \ch\ there is no known maximum limit of biological \ch\ surface flux. As such, it is unknown how much \ch\ can accumulate in the atmosphere of an \om-rich planet.

\section{Summary and conclusions \label{sec:conclusions}}
This study expands on previous work by considering the impacts of different amounts of atmospheric \ch\ on the \om-\oz\ relationship for an Earth-like planet. We find that the impact of varying \ch\ on the \om-\oz\ relationship is highly influenced by both the host star and the amount of \om\ in the atmosphere, in a manner similar to when \no\ is varied \citep{koza25}. Increasing \ch\ to 1000\% PAL in our high \ch\ models was found to have significantly more of an impact on \oz\ than when we decreased \ch\ to 10\% PAL with our low \ch\ models (Sec.~\ref{sec:CH4_chem}). The most striking result is that planets around the hottest hosts with high \om\ abundances ($>$50\% PAL) experienced the opposite response of \oz\ to \ch\ to all other models we explored. There are two main reasons for this effect. First, in high UV environments for models with plentiful \om\ and high \ch\ large amounts of stratospheric \hho\ were created, which resulted in cooler stratospheres that increased \oz\ production and slowed down its destruction, in a process similar to stratospheric cooling via CO$_2$ on modern Earth. Second, for these hotter hosts at high \om\ there was a boost in \oz-produced smog with the high \ch\ models, as the additional \hox\ created from \hho\ resulted in a higher \nox/\hox\ ratio than with modern levels of \ch. This allowed these ``\nox-saturated'' atmospheres to have faster smog mechanisms as it was more difficult for \nox\ to lock up the available \hox\ into reservoir species. For the hottest hosts at lower \om\ levels, as well as all models around cooler hosts, the high \ch\ models resulted in lower \oz\ abundances than compared with modern levels of \ch. This was primarily due to the increased efficiency of the \hox\ catalytic cycle in destroying \oz.

The largest absolute changes in \oz\ abundance due to variations in \ch\ occurred at higher \om\ levels, causing the amount of harmful UVC reaching the surface of our model planets to change significantly, especially for the high \ch\ models (Sec.~\ref{sec:UV_to_ground}). At 100\% PAL \om\ for the high \ch\ models the planets orbiting the Sun and M5V hosts experienced factors of 7.5$\times10^{-5}$ and 2.1$\times10^{4}$ times the amount of UVC surface flux, respectively, when compared to models with modern levels of \ch. When considering how this could impact future observations we looked at the 9.6~$\mu$m \oz\ feature in planetary emission spectra (Sect.~\ref{sec:emission_spectra}). We found again that the high \ch\ models had the most impact, with planets around all hosts being affected to some degree depending on the amount of \om\ in the atmosphere. Planets orbiting the G0V and Sun hosts with the high \ch\ models had the most change in the \oz\ feature dependent on the \om\ level, with the extra \ch\ sometimes causing a deeper or shallower feature.

These results further complicate the usage of \oz\ as a proxy for \om, but also provide additional guidance for future observations. We have now shown in this study that varying \ch\ impacts the \om-\oz\ relationship just as much as \no\ \citep{koza25}, but in different ways. There are many scenarios where high \ch\ could be increasing the \oz\ of an atmosphere, while high \no\ would be working at the same time to deplete that \oz. This shows that we would be required to think about variations of both species in order to use an \oz\ measurement to learn about the \om\ content of the atmosphere. Once again we gain no general rules about how \oz\ would be impacted when thinking about variations in \ch\ or \no, with the specific host star playing a highly influential role.

We find that untangling the impact of \ch\ and \no\ on the \om-\oz\ relationship might be more straightforward for cooler stars, especially around hosts like the M5V star used in this study as \no\ impacts on \oz\ are minimal, with \ch\ mainly impacting the atmosphere at higher \om\ abundances. Planets around hotter hosts like our G0V and Sun hosts have the potential to more significantly alter \oz\ measurements, as impacts from \ch\ and \no\ on \oz\ change significantly based on the \om\ content of the atmosphere. Careful modeling and retrieval studies exploring the parameter space of \om, \no, and \ch\ for Earth-like atmospheres will be necessary if we wish to glean information about the amount of \oz\ -- and therefore \om\ -- in a exoplanetary atmosphere when the time comes that we are able to perform such observations.

\begin{acknowledgements}
This project is funded by VILLUM FONDEN and all computing was performed on the HPC cluster at the Technical University of Denmark \citep{hpc}. Authors TK and LML acknowledge financial support from Severo Ochoa grant CEX2021-001131-S funded by MCIN/AEI/ 10.13039/501100011033. JMM acknowledges support from the Horizon Europe Guarantee Fund, grant EP/Z00330X/1. We thank the anonymous referee for their useful comments, which improved the clarity of our manuscript.
\end{acknowledgements}

\bibliographystyle{aa}
\bibliography{main.bib}{}

\begin{appendix}
\onecolumn
\section{Supplementary figures and tables \label{sec:appendix}}

\begin{figure*}[h!]
\centering
\includegraphics[scale=0.55]{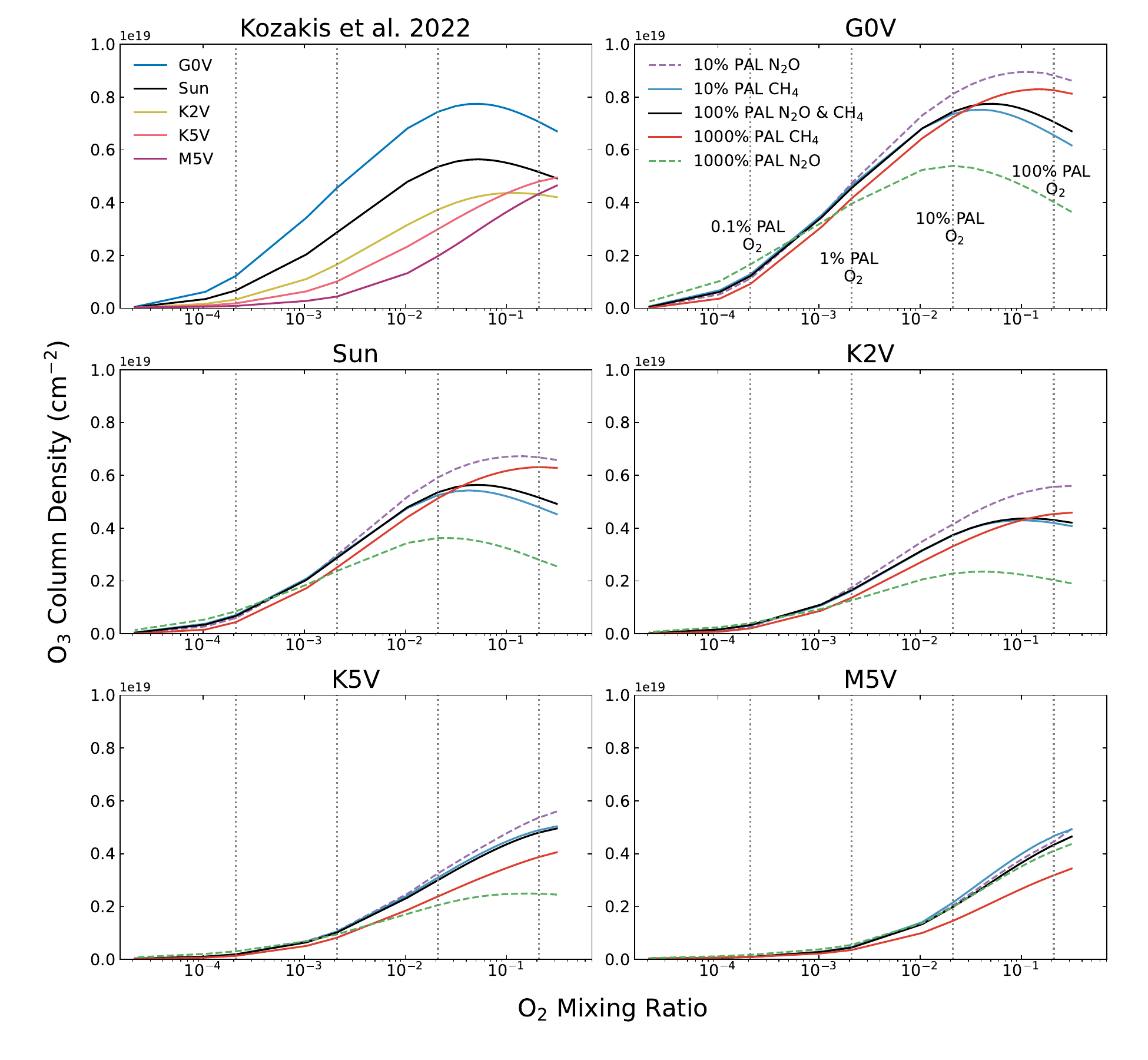}
    \caption{Relationships of \om-\oz\  for all host stars at all \om\ and \ch\ levels modeled, along with comparisons to varying levels of \no\ as modeled in \cite{koza25}. All plots share the same y-axis scale to facilitate comparisons. For planets around all hosts except the M5V there are larger variations in \oz\ when varying \no\ rather than \ch. Only the M5V-hosted planet exhibits stronger changes in \oz\ with \ch\ variations.
\label{fig:O2O3_relationship}}
\end{figure*}

\begin{figure*}[h!]
\centering
\includegraphics[scale=0.6]{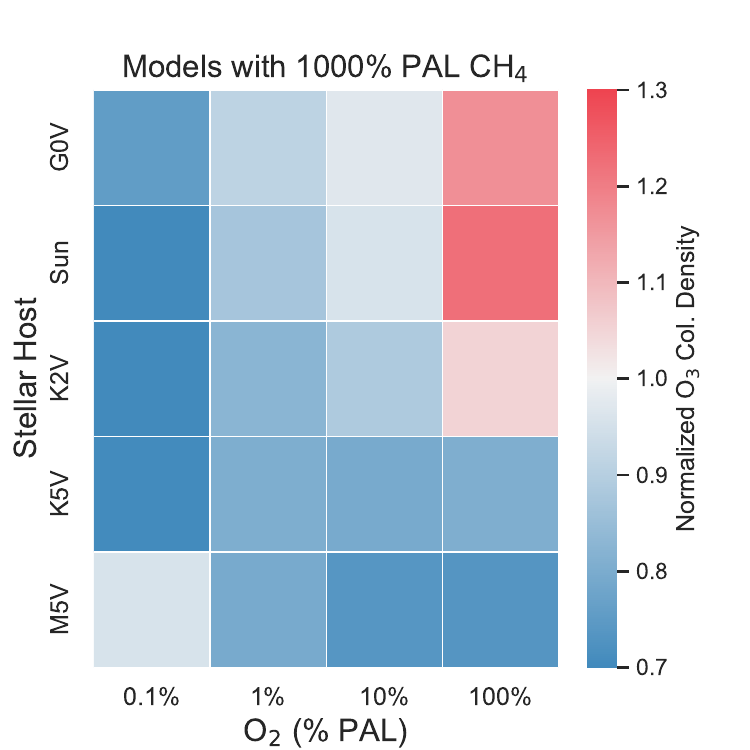}
\includegraphics[scale=0.6]{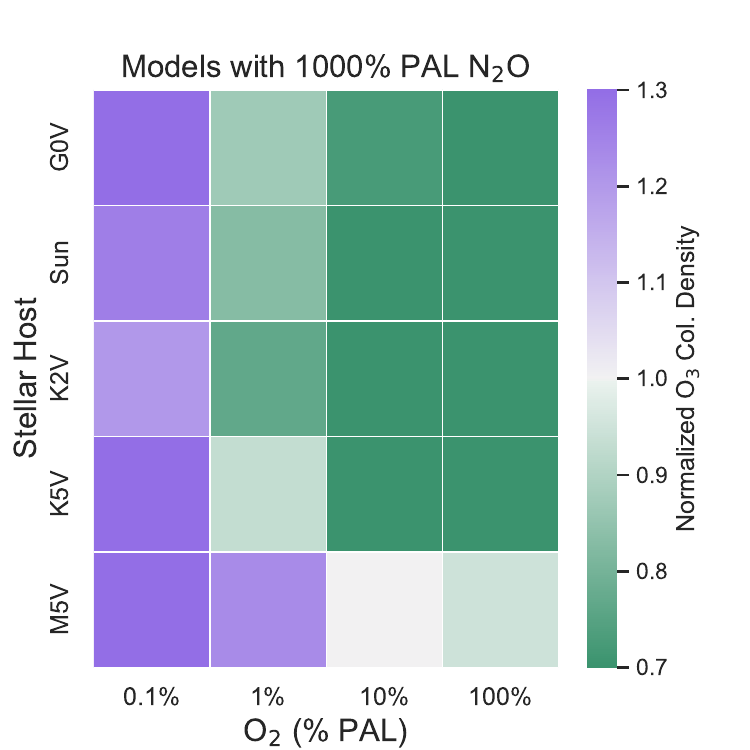}
    \caption{Abundances of \oz\ for both high \ch\ models from this study and high \no\ models from \cite{koza25} normalized to the amount of \oz\ with modern amounts of \ch\ and \no\ for all host stars at 0.1\%, 1\%, 10\%, and 100\% PAL \om. Both figures share the same color bar limits in order to facilitate comparisons. 
\label{fig:O2O3_grids_compare}}
\end{figure*}

\begin{figure*}
\centering
\includegraphics[scale=0.45]{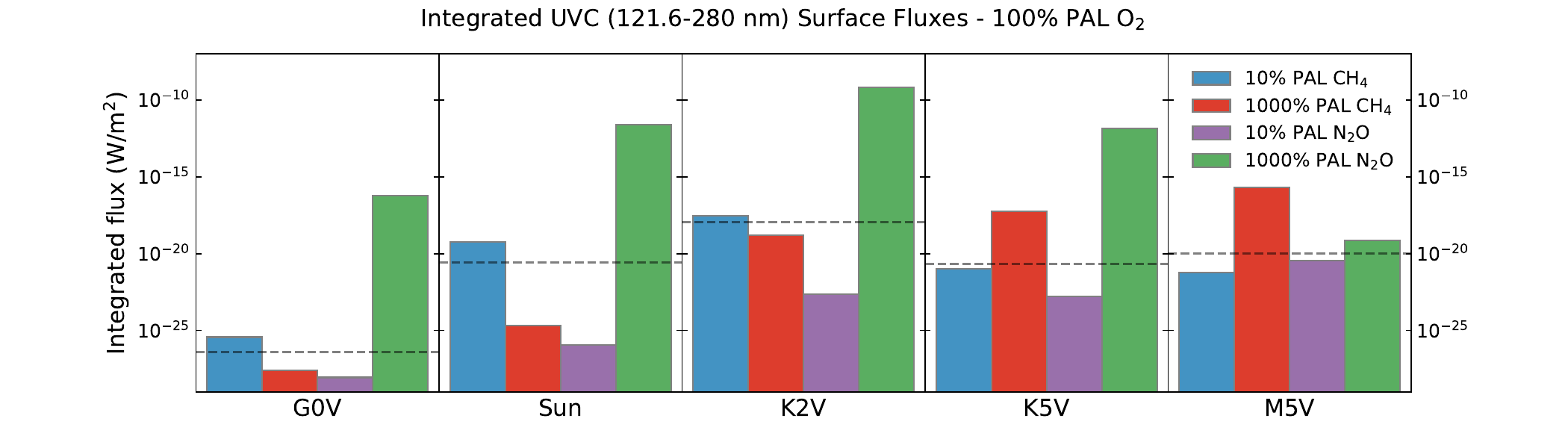}\\
\includegraphics[scale=0.45]{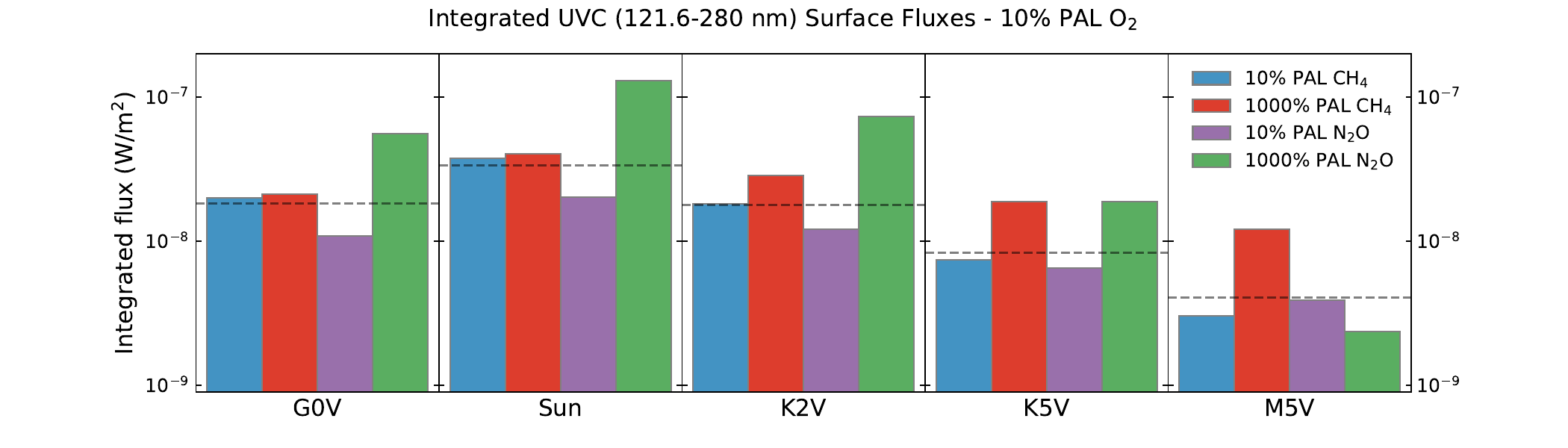}\\
\includegraphics[scale=0.45]{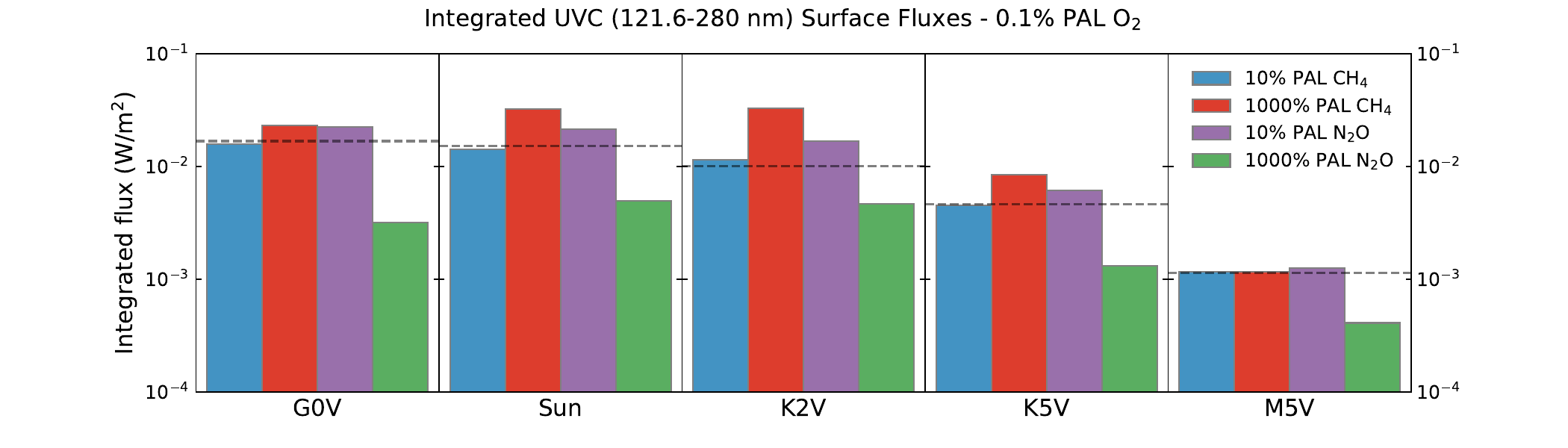}\\
    \caption{Comparisons of UVC surface flux for all hosts at 100\%, 10\% and 0.1\% PAL \om, as well as high and low \ch\ and \no\ variations from this study as well as \cite{koza25}. The dashed horizontal lines indicate the amount of \oz\ for models with modern levels of both \ch\ and \no. Overall variations in \no\ have a stronger impact on UVC surface flux, with larger changes when varying \ch\ present primarily at 100\% PAL \om.}
\label{fig:UVbarchart}
\end{figure*}

{\singlespace
\begin{table*}[h!]
\centering
\footnotesize
\caption{UV integrated fluxes \label{tab:UV_all}}
\begin{tabular}{crrr|rr|rr}
Spectral & O$_2$ MR & TOA flux & \cite{koza22}& \multicolumn{4}{c}{Surface Flux Normalized to \cite{koza22}}\\
\cline{5-8}
Type & (\% PAL) & (W/m$^2$) & Surface Flux (W/m$^2$)  & 10\% PAL \no\ & 1000\% PAL \no\ & 10\% PAL \ch\ & 1000\% PAL \ch\ \\
\hline
\hline 
\multicolumn{8}{c}{UVB Fluxes (280 - 315 nm)}\\
\hline
G0V 	 & 100	 & 22.35	 & 1.5e+00	 & 0.72	 & 1.84	 & 1.10	 & 0.81 \\
G0V 	 & 10	 & 22.35	 & 1.4e+00	 & 0.88	 & 1.47	 & 1.01	 & 1.05 \\
G0V 	 & 1	 & 22.35	 & 2.4e+00	 & 0.97	 & 1.13	 & 0.98	 & 1.09 \\
G0V 	 & 0.1	 & 22.35	 & 5.9e+00	 & 1.05	 & 0.84	 & 0.98	 & 1.14 \\
\hline
Sun 	 & 100	 & 16.18	 & 1.6e+00	 & 0.74	 & 1.69	 & 1.07	 & 0.80 \\
Sun 	 & 10	 & 16.18	 & 1.6e+00	 & 0.89	 & 1.43	 & 1.02	 & 1.05 \\
Sun 	 & 1	 & 16.18	 & 2.7e+00	 & 0.98	 & 1.13	 & 0.99	 & 1.10 \\
Sun 	 & 0.1	 & 16.18	 & 5.6e+00	 & 1.05	 & 0.90	 & 0.98	 & 1.15 \\
\hline
K2V 	 & 100	 & 4.8	 & 6.8e-01	 & 0.79	 & 1.67	 & 1.02	 & 0.95 \\
K2V 	 & 10	 & 4.8	 & 7.4e-01	 & 0.92	 & 1.39	 & 1.00	 & 1.09 \\
K2V 	 & 1	 & 4.8	 & 1.2e+00	 & 0.97	 & 1.13	 & 1.01	 & 1.09 \\
K2V 	 & 0.1	 & 4.8	 & 2.2e+00	 & 1.04	 & 0.95	 & 1.01	 & 1.10 \\
\hline
K5V 	 & 100	 & 0.68	 & 9.8e-02	 & 0.90	 & 1.60	 & 0.98	 & 1.19 \\
K5V 	 & 10	 & 0.68	 & 1.4e-01	 & 0.95	 & 1.24	 & 0.98	 & 1.15 \\
K5V 	 & 1	 & 0.68	 & 2.3e-01	 & 0.99	 & 1.02	 & 0.99	 & 1.07 \\
K5V 	 & 0.1	 & 0.68	 & 3.4e-01	 & 1.02	 & 0.92	 & 1.00	 & 1.05 \\
\hline
M5V 	 & 100	 & 3.5e-02	 & 6.5e-03	 & 0.98	 & 1.04	 & 0.94	 & 1.24 \\
M5V 	 & 10	 & 3.5e-02	 & 1.0e-02	 & 0.99	 & 1.00	 & 0.96	 & 1.13 \\
M5V 	 & 1	 & 3.5e-02	 & 1.6e-02	 & 1.00	 & 0.96	 & 0.99	 & 1.04 \\
M5V 	 & 0.1	 & 3.5e-02	 & 2.0e-02	 & 1.01	 & 0.94	 & 1.00	 & 1.00 \\
\hline
\hline
\multicolumn{8}{c}{UVC Fluxes (121.6 - 280 nm)} \\
\hline
G0V 	 & 100	 & 11.2	 & 3.8e-27	 & 2.3e-02	 & 1.5e+10	 & 9.7e+00	 & 6.8e-02 \\
G0V 	 & 10	 & 11.2	 & 1.8e-08	 & 5.9e-01	 & 3.1e+00	 & 1.1e+00	 & 1.2e+00 \\
G0V 	 & 1	 & 11.2	 & 1.7e-04	 & 1.0e+00	 & 5.2e-01	 & 9.2e-01	 & 1.4e+00 \\
G0V 	 & 0.1	 & 11.2	 & 1.7e-02	 & 1.3e+00	 & 1.9e-01	 & 9.4e-01	 & 1.4e+00 \\
\hline
Sun 	 & 100	 & 6.7	 & 2.8e-21	 & 4.1e-06	 & 8.8e+08	 & 2.2e+01	 & 7.5e-05 \\
Sun 	 & 10	 & 6.7	 & 3.3e-08	 & 6.0e-01	 & 3.9e+00	 & 1.1e+00	 & 1.2e+00 \\
Sun 	 & 1	 & 6.7	 & 2.3e-04	 & 1.0e+00	 & 7.2e-01	 & 9.4e-01	 & 1.5e+00 \\
Sun 	 & 0.1	 & 6.7	 & 1.5e-02	 & 1.4e+00	 & 3.2e-01	 & 9.3e-01	 & 2.1e+00 \\
\hline
K2V 	 & 100	 & 1.4	 & 1.1e-18	 & 2.0e-05	 & 6.0e+08	 & 2.6e+00	 & 1.5e-01 \\
K2V 	 & 10	 & 1.4	 & 1.8e-08	 & 6.8e-01	 & 4.1e+00	 & 1.0e+00	 & 1.6e+00 \\
K2V 	 & 1	 & 1.4	 & 1.0e-04	 & 9.8e-01	 & 8.4e-01	 & 1.0e+00	 & 1.4e+00 \\
K2V 	 & 0.1	 & 1.4	 & 1.0e-02	 & 1.7e+00	 & 4.6e-01	 & 1.1e+00	 & 3.2e+00 \\
\hline
K5V 	 & 100	 & 0.16	 & 2.1e-21	 & 7.8e-03	 & 6.6e+08	 & 4.6e-01	 & 2.7e+03 \\
K5V 	 & 10	 & 0.16	 & 8.4e-09	 & 7.8e-01	 & 2.3e+00	 & 8.8e-01	 & 2.3e+00 \\
K5V 	 & 1	 & 0.16	 & 5.4e-05	 & 1.0e+00	 & 5.7e-01	 & 9.4e-01	 & 1.5e+00 \\
K5V 	 & 0.1	 & 0.16	 & 4.7e-03	 & 1.3e+00	 & 2.8e-01	 & 9.7e-01	 & 1.8e+00 \\
\hline
M5V 	 & 100	 & 2.7e-02	 & 9.8e-21	 & 3.7e-01	 & 7.4e+00	 & 5.8e-02	 & 2.1e+04 \\
M5V 	 & 10	 & 2.7e-02	 & 4.1e-09	 & 9.7e-01	 & 5.9e-01	 & 7.5e-01	 & 3.0e+00 \\
M5V 	 & 1	 & 2.7e-02	 & 3.8e-05	 & 1.0e+00	 & 4.0e-01	 & 8.8e-01	 & 1.8e+00 \\
M5V 	 & 0.1	 & 2.7e-02	 & 1.1e-03	 & 1.1e+00	 & 3.6e-01	 & 1.0e+00	 & 1.0e+00 \\
\hline
\hline
\end{tabular}
\vspace{-0.2cm}
\tablefoot{
Abbreviations: MR = mixing ratio; PAL = present atmospheric level; TOA =  top of atmosphere}\end{table*}
}

\begin{figure*}
\centering
\includegraphics[scale=0.5]{allO2_emission_O3_T_panels_onlyemission_paper2.pdf}
\hspace{-1.2cm}
\includegraphics[scale=0.5]{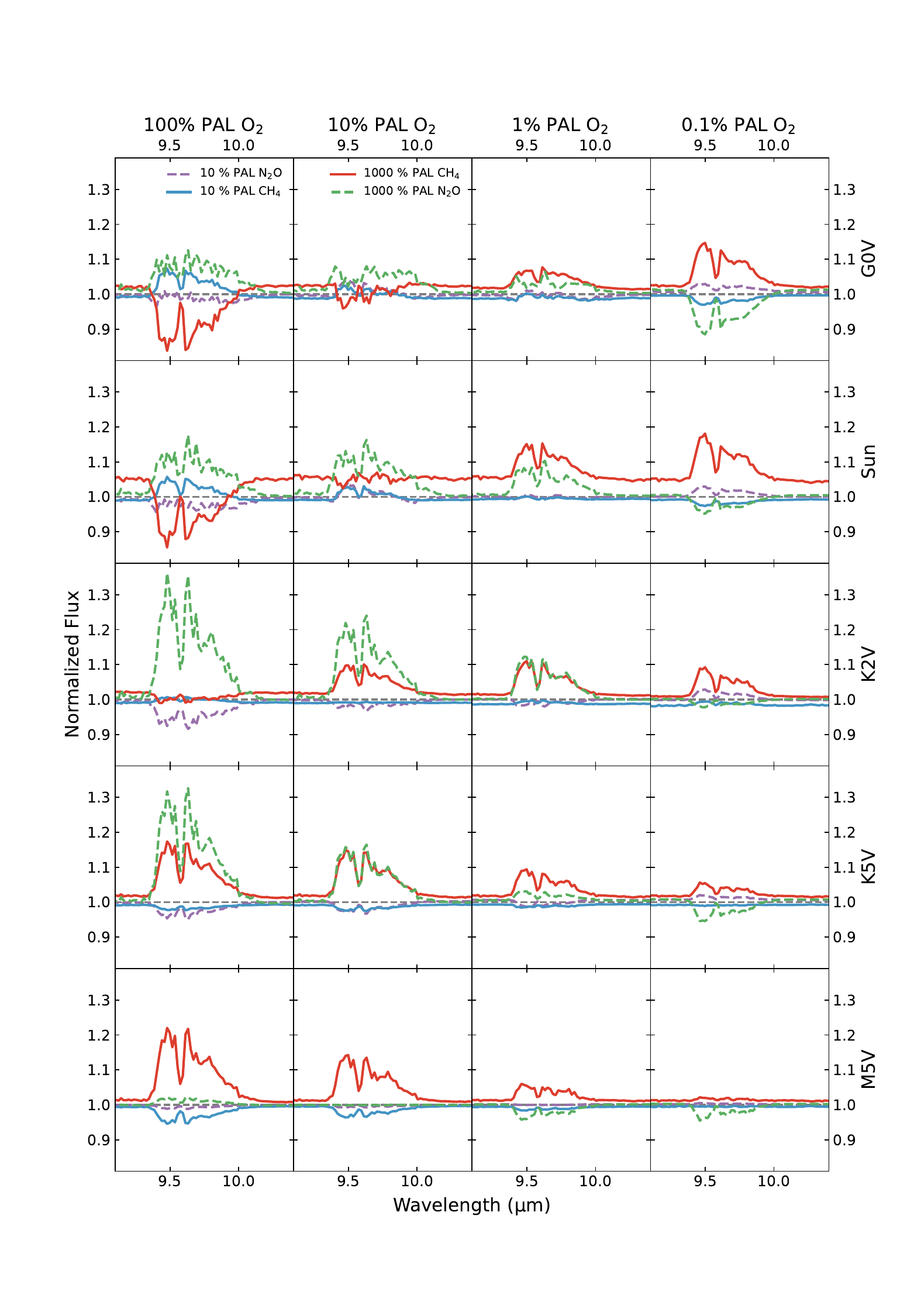}
\vspace{-1.5cm}
\caption{Comparisons of 9.6~$\mu$m \oz\ emission spectra features from \cite{koza22} with modern levels of \ch\ and \no\ (left) and \oz\ features from varying \ch\ and \no\ models normalized to modern amounts of \ch\ and \no\ (right). Y-axis limits for all normalized features are the same to allow for comparison between different \om\ levels and host stars. For both variations in \ch\ and \no\ changes in the \oz\ feature were primarily due to differences in \oz\ abundance rather than changes in the atmospheric temperature profiles. The exception being for the \ch\ models for the hotter stars at 100\% PAL \om, which experienced stratospheric temperature changes depending on the amount of \hho\ produced from \ch.
\label{fig:emission_N2O_CH4}
}
\end{figure*}

\end{appendix}

\end{document}